# Slender body theory for particles with non-circular cross-sections with application to particle dynamics in shear flows

## Neeraj S. Borker[1] and Donald L. Koch[2]

[1]Sibley School of Mechanical and Aerospace Engineering, Cornell University, Ithaca, NY 14853, USA

[2]Robert Frederick Smith School of Chemical and Biomolecular Engineering, Cornell University, Ithaca, NY 14853, USA

## Abstract

This paper presents a theory to obtain the force per unit length acting on a slender filament with a non-circular cross-section moving in a fluid at low Reynolds number. Using a regular perturbation of the inner solution, we show that the force per unit length has $O(1/\ln(2A)) + O(\alpha/\ln^2(2A))$ contributions driven by the relative motion of the particle and the local fluid velocity and an $O(\alpha/(\ln(2A)A))$ contribution driven by the gradient in the imposed fluid velocity. Here, the aspect ratio ($A = l/a_0$) is defined as the ratio of the particle size ($l$) to the cross-sectional dimension ($a_0$); and $\alpha$ is the amplitude of the non-circular perturbation. Using thought experiments, we show that two-lobed and three-lobed cross-sections affect the response to relative motion and velocity gradients, respectively. A two-dimensional Stokes flow calculation is used to extend the perturbation analysis to cross-sections that deviate significantly from a circle (i.e., $\alpha \sim O(1)$). We demonstrate the ability of our method to accurately compute the resistance to translation and rotation of a slender triaxial ellipsoid. Furthermore, we illustrate novel dynamics of straight rods in a simple shear flow that translate and rotate quasi-periodically if they have two-lobed cross-section; and rotate chaotically and translate diffusively if they have a combination of two- and three-lobed cross-sections. Finally, we show the remarkable ability of our theory to accurately predict the motion of rings, retaining great accuracy for moderate aspect ratios ($\sim 10$) and cross-sections that deviate significantly from a circle, thereby making our theory a computationally inexpensive alternative to other Stokes flow solvers.

## 1. Introduction

Slender geometries are quite common in the realm of low Reynolds number fluid mechanics. Many microorganisms such as E. coli (Berg and Anderson 1973), Chlamydomonas (Bray 2000), Paramecium (Brennen and Winet 1977) or human spermatozoa (Suarez and Pacey 2006) use slender appendages to navigate through viscous fluid environments. The orientation distribution of plankton, some of which have elongated geometries, in the ocean is known to strongly influence the propagation of light in the deeper levels which can significantly influence the global carbon fixation (Guasto, Rusconi and Stocker 2012). Slender particles such as fibers are used to make fiber-reinforced composites that have enhanced tensile strength and increased anisotropic thermal conductivity (Tekce, Kumlutas & Tayman 2007). Slender body theory (Cox 1970; Batchelor 1970; Keller and Rubinow 1976; and Johnson 1980) provides a computationally inexpensive route to study the dynamics of slender particles in highly viscous flows. However, the influence of the force per unit length generated by the gradient in the imposed fluid velocity has not been previously embedded in this theory.

The force per unit length experienced by a slender filament with a circular cross-section at low Reynolds number, to leading order, is qualitatively similar to the viscous drag experienced by a long circular cylinder due to its relative motion with the local fluid velocity. Here, the existence of an additional force per unit length driven by the gradient in the imposed fluid velocity and its dependence on the shape of the cross-section is elucidated. In this study, the nature of the velocity disturbance is understood using a regular perturbation of the inner solution of the slender body theory (SBT) formulation. An integral equation is derived for the force per unit length experienced by the slender filament along with its dependence on the cross-sectional geometry. The force per unit length due to the non-circularity of the cross-section, has two components, one which is driven by the gradient in the imposed fluid velocity affecting three-lobed cross-sections (figure 1(c)) and another driven by the motion of the particle relative to the fluid affecting two-lobed cross-sections (figure 1 (b)), a component first presented by Batchelor (1970). The additional force per unit length driven by the gradient in the fluid velocity is extremely important in the SBT formulation especially for computing the dynamics and rheology of thin particles in a linear flow field. A numerical calculation of the inner solution in a two-dimensional domain is proposed to extend our SBT formulation to particles with a general cross-sectional shape.

The current theory accurately predicts the resistance to rotation and translation of slender triaxial ellipsoids retaining accuracy even when the cross-section has a high aspect ratio (section 4). In section 5, straight particles with three-lobed cross-sections, illustrated in figure 1 (c), are shown to rotate and translate quasi-periodically in a simple shear flow (SSF) because of the force per unit length driven by the imposed velocity gradient. In contrast, a straight particle with a combination of a two- and a three-lobed cross-section, shown in figure 1 (d), can diffuse in space while rotating chaotically. This work allows for the inclusion of these dynamics which can potentially impact the rheology of a suspension of straight particles. In section 6, our theory is utilized to predict the dynamics of rings with non-circular cross-section, and the results remain accurate for cross-sections that deviate significantly from a circle or have aspect ratios as low as 10.

Slender body theory (SBT) allows for an approximate solution of the governing equation modeling a physical phenomenon that is affected by the presence of bodies which are long and thin. The governing equations of highly viscous flows (Stokes flow), potential flows and heat transfer have been solved using the SBT formulation for many applications. Potential flow problems that include animal locomotion (Lighthill (1960, 1971)), force on airship hulls (Munk, 1924), force on wings (Jones, 1946) and ship hydrodynamics (Newman, (1964, 1970)) have been solved using SBT. Steady state heat transfer in composites (Rocha and Acrivos, 1973 (a) & (b); Chen and Acrivos; Acrivos and Shaqfeh, 1988; Shaqfeh 1988; Fredrickson and Shaqfeh, 1989; Mackaplow and Shaqfeh. 1994;) and transient heat transfer in ground-source heat pumps (Beckers, Koch and Tester, 2014) are a few applications of SBT for steady state and transient heat conduction respectively. In the realm of highly viscous flows, problems involving flagellar hydrodynamics (Johnson and Brokaw, 1979), structure and rheology of fiber suspensions (Mackaplow and Shaqfeh. (1996, 1998); Rahnama, Koch, Iso and Cohen (1993); Rahnama, Koch and Shaqfeh, (1995)) and separation of racemic mixtures of screw shaped particles (Kim and Rae, 1991) have been solved using existing SBT techniques for Stokes flow (Cox (1970); Batchelor (1970); Keller and Rubinow (1976); and Johnson (1980)).

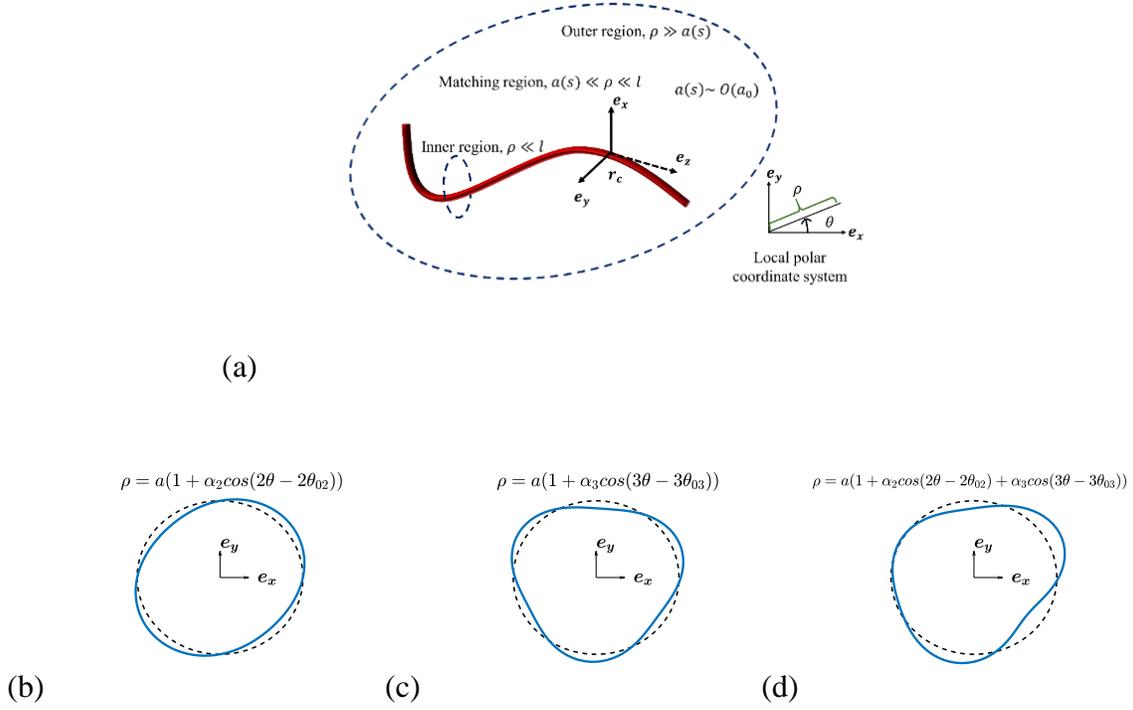

(a)

(b)                              (c)                              (d)

Figure 1. Local coordinate systems in the particle reference frame and the perturbed cross-sectional shapes studied here. (a) Local coordinate system for a general curved body. $\boldsymbol{e_z}$ is along the tangent to the filament axis, $\boldsymbol{e_x}$ is along the normal and $\boldsymbol{e_y}$ is pointed along the binormal to the centerline of the slender body ($\boldsymbol{r_c}$). (b) Schematic of a two-lobed shape (S-I), given by $\rho = a\,(1 + \alpha_2\,cos(2\theta - 2\theta_{02}))$. (c) Schematic of a three-lobed shape (S-II), given by $\rho = a\,(1 + \alpha_3\,cos(3\theta - 3\theta_{03}))$. (d) Schematic of a shape, which is a linear combination of the two and three lobed perturbations, given by $\rho = a\,(1 + \alpha_2\,cos(2\theta - 2\theta_{02}) + \alpha_3\,cos(3\theta - 3\theta_{03}))$ (S-III shape). "$a$" the radius of the equivalent circle is allowed to vary along the centerline of the filament.

The basic idea in SBT is to obtain the strength of a line of singularities placed along the centerline of the slender filament that approximates the field of interest around the filament far away from the cross-sectional surface, termed as the outer region, i.e, $a \ll \rho$. Here $\rho$ is the radial distance from the centerline of the slender filament, and "$a$" is a measure of the cross-sectional size of the particle at a certain location along the centerline of the slender body as shown in figure 1 (a). The singularity for a potential flow problem is a point source of mass, for a heat transfer problem a point source of heat and for a Stokes flow problem a point force. The strength of the singularities is found by matching the field approximated in the outer region, termed as the outer solution, to a field obtained from the inner region (i.e, $\rho \ll l$, where l is the length of the slender filament). Placing higher order singularities along the centerline of the slender filament gives a better estimate of the field of interest. For a Stokes flow problem, which is the main topic of discussion in this paper, these singularities would include doublets, rotlets, sources, stresslets and quadrupoles (Johnson, 1980).

In Stokes flow, the force per unit length exerted by the body on the fluid (i.e. the strength of the singularity) is derived at each point along the centerline of the slender body in terms of the motion of the particle and the imposed fluid velocity. The length, "$l$", an appropriate velocity and the fluid viscosity are used to non-dimensionalize all variables throughout this paper unless mentioned otherwise. The force per unit length exerted by a slender body on the fluid, which is due to relative motion of the particle and fluid, can be expressed as a series in $\epsilon = 1/\ln(2A)$ (Cox, 1970 and Batchelor, 1970), where $A = l/(a_0)$ is the aspect ratio of the particle and $a_0$ is the characteristic value of $a(s)$. The first term in this series arises only due to the local relative velocity between the particle and the fluid and is O($\epsilon$) (Cox, 1970 and Batchelor, 1970). Cox (1970) was the first to demonstrate that the force per unit length is affected at O($\epsilon^2$) by the centerline curvature (see equations (6.2) and (6.3) of his paper). Keller and Rubinow (1976) gave an integral equation for the force per unit length which can be iteratively solved to obtain higher order corrections to the force per unit length with errors of O($\epsilon^N$), where N is an integer such that $N \geq 2$. Johnson (1980) produced an integral equation for the force per unit length with errors of O($1/A^2$) which also included the effects of the ends of the slender body (equation (19) of his paper). Johnson (1980) also described a method to incorporate the effect of centerline curvature in the inner solution (equation (31) of his paper).

Batchelor (1970) showed that the cross-sectional shape of the particle first affects the force per unit length at O($\epsilon^2$), the same order of importance as the centerline curvature of the body. Batchelor (1970) described how the force per unit length for an arbitrary cross-section can be found by solving for the longitudinal velocity field due to an equivalent circle of a certain radius, and the transverse velocity field due to an equivalent ellipse of certain dimensions and orientation. Batchelor (1954) provided the method to obtain the equivalent circular cylinder by solving the harmonic equation for the flow along the longitudinal direction. Batchelor (1970) described the procedure to obtain an equivalent ellipse by solving the biharmonic equation for the velocity field in the transverse plane around the cross-sectional shape in question.

Cox (1971) was the first to account for the effect of the gradient in the imposed velocity of a linear flow field on the force and torque acting on a slender cylinder with tapered ends, which was oriented such that the imposed velocity field at the particle centerline is exactly zero. Cox placed a force and dipole per unit length along the centerline of the body to approximate the velocity field in the outer region, while Keller and Rubinow (1976) employed a stresslet and rotlet per unit length for the same purpose. Singh, Koch, Subramanian and Stroock (2014) discuss the equivalence of these two formulations by showing that the variation of the force per unit length leads to a net stresslet and a rotlet on a slice of the particle. The rotlet per unit length can lead to a finite torque on the particle whereas a variable stresslet per unit length, which occurs when the particle cross-section varies along the longitudinal direction, can lead to a finite force acting on the particle of $O(\epsilon^2/A^2)$. A straight circular cylinder with a constant cross-section cannot experience a force per unit length at this orientation if the cross-section is circular. However, the force per unit length due to the velocity gradient can be non-zero if the cross-section is non-circular. This force per unit length, which will be $O(\epsilon/A)$ as explained in section 3, can lead to a net force of $O(\epsilon/A)$ acting on the particle.

In this paper, the additional $O(\epsilon/A)$ contribution to the force per unit length in addition to Batchelor's correction of $O(\epsilon^2)$ is derived in section 3. This calculation is especially important when considering motion of force and torque-free slender particles in SSF. In such scenarios, Batchelor's correction tends to zero for certain particle orientations, while Cox's (1971) correction predicts a much smaller effect on the particle rotation rate. Thus, both previous theories lead to an incorrect qualitative behavior for particle geometries such as straight cylinders with elliptic cross-sections (Yarin, Gottlieb and Roisman 1997) or rings with 3-lobed cross-sections (Singh, Koch and Stroock, 2013) in SSF.

## 2. SBT solution for circular cross-section

In this section, the velocity disturbance created by a slender particle with a circular cross-section, when placed in a fluid moving with a velocity $\boldsymbol{u}_\infty$ in the absence of the particle is described. This calculation will be used to obtain the effect of the cross-sectional shape in section 3. The slenderness parameter or the aspect ratio ($A = l/(a_0)$), is defined as the ratio of the half-length of the filament ($l$) and a measure of the radius of the cross-section ($a_0$). The radius of the cross-section, $a \sim O(a_0)$, is allowed to vary along the longitudinal direction of the slender body. The position vector is denoted by $\boldsymbol{r}$ and $\boldsymbol{r}_c$ denotes the position of the centerline of the slender filament. A local coordinate system ($\boldsymbol{e}_x, \boldsymbol{e}_y, \boldsymbol{e}_z$) is chosen based on the tangent ($\boldsymbol{e}_z$), normal ($\boldsymbol{e}_x$) and binormal ($\boldsymbol{e}_y$) to the centerline of the slender body, as shown in figure 1 (a), and is mathematically given by

$$\boldsymbol{e}_z = \frac{\partial \boldsymbol{r}_c}{\partial s}, \boldsymbol{e}_x = \frac{1}{\kappa}\frac{\partial^2 \boldsymbol{r}_c}{\partial s^2}, \ \boldsymbol{e}_y = \boldsymbol{e}_z \times \boldsymbol{e}_x, \tag{2.1}$$

where "$s$" denotes the arc length along the centerline and $\kappa$ is the local curvature of the body centerline. A local polar coordinate system ($\rho$-$\theta$), as shown in figure 1 (a), is defined in the $\boldsymbol{e}_x - \boldsymbol{e}_y$ plane, where $\theta$ is measured from $\boldsymbol{e}_x$ and $\rho$ is the normal distance from the centerline. The far-field fluid velocity is denoted by $\boldsymbol{u}_\infty$, which is allowed to vary with $\boldsymbol{r}$. The velocity on the particle surface ($\boldsymbol{r} = \boldsymbol{r}_s$) is given by

$$\boldsymbol{u}(\boldsymbol{r} = \boldsymbol{r}_s) = \boldsymbol{U} + \boldsymbol{\omega} \times \boldsymbol{r}_s = (\boldsymbol{U} + \boldsymbol{\omega} \times \boldsymbol{r}_c) + \boldsymbol{\omega} \times (\boldsymbol{r}_s - \boldsymbol{r}_c), \tag{2.2}$$

where $\boldsymbol{U}$ and $\boldsymbol{\omega}$ are the particle velocity at the origin ($\boldsymbol{r} = \boldsymbol{0}$) and angular velocity of the particle respectively. These definitions are valid for any cross-sectional shape. The rest of this section describes the slender body theory solution for a circular cross-section of radius "a". The force per unit length and all the unknown constants in the inner solution, which are required for the perturbation analysis in section 3, are obtained in this section.

*2.1 Velocity field in the inner region ($\rho \ll 1$)*

Any curved slender body with $O(1)$ curvature appears locally as a straight infinite cylinder to a first approximation. The velocity field in the inner region is obtained by assuming flow over an infinite cylinder. Thus, the flow along and transverse to the cylinder is solved separately. Any coupling between these flows arises due to curvature and finite aspect ratio of the particle and leads to algebraic $O(1/A^2)$ corrections to the velocity disturbance (Johnson 1980) which are not discussed here. The functional form of the flow in the transverse plane is obtained by solving a two-dimensional Stokes flow problem that satisfies the no-slip condition on the particle surface. The far field boundary condition is applied later while asymptotically matching the velocity fields from the inner and outer regions to obtain any unknowns. The solution to the biharmonic equation ($\nabla^4 \psi = 0$) is used to obtain the functional form of the velocity field in the plane transverse to the slender dimension. The two-dimensional velocity field is obtained from the definition of the stream function, i.e., $u_\rho = \rho^{-1} d\psi/d\theta$ and $u_\theta = -d\psi/d\rho$. The solution around a circular cross-section of radius "a" in terms of the stream-function ($\psi$) in polar coordinates is given by

$$\frac{\psi}{a} = \left(\frac{\tilde{\psi}}{a}\right) + \left[B\cos(\theta) + \hat{B}\sin(\theta)\right]\left[\left(\frac{\rho}{a}\right) - \left(\frac{\rho}{a}\right)^{-1} - 2\ln\left(\frac{\rho}{a}\right)\left(\frac{\rho}{a}\right)\right] + \left[(\boldsymbol{U} + \boldsymbol{\omega} \times \boldsymbol{r_c}) \cdot (\boldsymbol{e_x}\sin(\theta) - \boldsymbol{e_y}\cos(\theta))\right]\left[\left(\frac{\rho}{a}\right)^{-1} + 2\frac{\rho}{a}\ln\left(\frac{\rho}{a}\right)\right] - (\boldsymbol{\omega} \cdot \boldsymbol{e_z})a\ln\left(\frac{\rho}{a}\right), \tag{2.3}$$

where $\tilde{\psi}$ is the stream function that approaches the stream function of the imposed flow field for $\rho \gg a$, while satisfying the zero-velocity boundary condition on the particle surface. B and $\hat{B}$ are obtained by matching this velocity field to the one from the outer region and depend on the force per unit length acting on the slender body, the imposed flow field and the particle velocities ($\boldsymbol{U}$ and $\boldsymbol{\omega}$). The velocity in polar coordinates in the inner region is given by

$$u_\rho = \frac{1}{\rho}\frac{\partial \psi}{\partial \theta} = \frac{1}{\rho}\frac{\partial \tilde{\psi}}{\partial \theta} + \left[-B\sin(\theta) + \hat{B}\cos(\theta)\right]\left[1 - \left(\frac{\rho}{a}\right)^{-2} - 2\ln\left(\frac{\rho}{a}\right)\right] + \left[(\boldsymbol{U} + \boldsymbol{\omega} \times \boldsymbol{r_c}) \cdot (\boldsymbol{e_x}\cos(\theta) + \boldsymbol{e_y}\sin(\theta))\right]\left[\left(\frac{\rho}{a}\right)^{-2} + 2\ln\left(\frac{\rho}{a}\right)\right], \tag{2.4}$$

$$u_\theta = -\frac{\partial \psi}{\partial \rho} = -\frac{\partial \tilde{\psi}}{\partial \rho} + \left[ B \cos(\theta) + \hat{B} \sin(\theta) \right] \left[ 1 - \left( \frac{\rho}{a} \right)^{-2} + 2 \ln \left( \frac{\rho}{a} \right) \right] + \left[ (\boldsymbol{U} + \boldsymbol{\omega} \times \boldsymbol{r_c}) \cdot \right.$$

$$\left. (\boldsymbol{e_x} \sin(\theta) - \boldsymbol{e_y} \cos(\theta)) \right] \left[ \left( \frac{\rho}{a} \right)^{-2} - 2 - 2 \ln \left( \frac{\rho}{a} \right) \right] + \boldsymbol{\omega} \cdot \boldsymbol{e_z} a \left( \frac{\rho}{a} \right)^{-1}. \qquad (2.5)$$

The velocity along the longitudinal direction is obtained by assuming negligible change in pressure along the longitudinal direction (i.e. $\nabla^2 u_z = 0$). The velocity field along a slender filament with circular cross-section is given by

$$u_z = \tilde{u}_z + E \ln \left( \frac{\rho}{a} \right) + (\boldsymbol{U} + \boldsymbol{\omega} \times \boldsymbol{r_c}) \cdot \boldsymbol{e_z} + \left( \frac{\rho}{a} \right)^{-1} (\boldsymbol{\omega} \cdot \boldsymbol{e_x} a \sin(\theta) - \boldsymbol{\omega} \cdot \boldsymbol{e_y} a \cos(\theta)), \qquad (2.6)$$

where $\tilde{u}_z$ is the velocity field that approaches $\boldsymbol{u}_\infty \cdot \boldsymbol{e_z}$ for $\rho/a \gg 1$, and equals zero on the particle surface, $E$ is obtained by matching the inner region velocity field to the outer solution and depends on the imposed flow field, the particle velocities ($\boldsymbol{U}$ and $\boldsymbol{\omega}$) and the force per unit length on the filament. The overall error in equations (2.4) - (2.6) is the larger of $O(1/A^2)$ and the order of errors in $B, \hat{B}$ and $E$, determined by matching the inner velocity field to the outer solution.

## 2.2 Velocity field in the outer region $\left( \rho \gg a(s) \right)$

In the outer region, the velocity disturbance due to a slender filament is approximately captured by a suitable choice of Stokeslet distribution along the particle centerline ($\boldsymbol{r_c}$) (Cox, 1970). A Stokeslet is a point force solution to the Stokes equations. The velocity disturbance created by a line of force per unit length is given by

$$\mathbf{u}(\boldsymbol{r}) \approx \boldsymbol{u}_\infty(\boldsymbol{r}) + \frac{1}{8\pi} \int_{\boldsymbol{r_c}} ds' \left( \frac{\boldsymbol{I}}{|\boldsymbol{r} - \boldsymbol{r'}|} + \frac{(\boldsymbol{r} - \boldsymbol{r'})(\boldsymbol{r} - \boldsymbol{r'})}{|\boldsymbol{r} - \boldsymbol{r'}|^3} \right) \cdot \boldsymbol{f}(\boldsymbol{r'}), \qquad (2.7)$$

where $\boldsymbol{r}$ is the point at which the velocity is evaluated, $\boldsymbol{r'}$ takes all values along the centerline and $ds'$ is the elemental length along the centerline of the slender body. Equation (2.7) has errors of $O(1/A^2)$. As $\boldsymbol{r'} \to \boldsymbol{r}$, the integrand in equation (2.7) diverges as $\ln(\rho)$. This diverging part of the integral in equation (2.7) is separated from the rest of the integral as shown by Keller and Rubinow (1976) and the resulting equation for $\rho \ll 1$ is given by

$$u(r) \approx u_\infty(r) - \frac{1}{4\pi}(I + e_z e_z) \cdot f(r) \left[\ln\left(\frac{\rho}{2}\right) - \ln(\sqrt{1-s^2})\right] - \frac{1}{4\pi} f \cdot e_z e_z + \frac{1}{4\pi} f \cdot e_\rho e_\rho +$$

$$\frac{1}{8\pi} \int_{r_c} ds' \left[\left(\frac{I}{|r_c(s)-r_c(s')|} + \frac{(r_c(s)-r_c(s'))(r_c(s)-r_c(s'))}{|r_c(s)-r_c(s')|^3}\right) \cdot f(r') - \left(\frac{I}{|s-s'|} + \frac{e_z e_z}{|s-s'|}\right) \cdot f(r) \right], \quad (2.8)$$

where $e_\rho$ is the radial vector in the $e_x - e_y$ plane. The integral on the right-hand-side of equation (2.8) is shown to have a finite limit by Keller and Rubinow (1976). The $\ln(\rho)$ term in equation (2.8), matches the $\ln(\rho)$ of the inner solution in equations (2.4)- (2.6) and also corresponds to the velocity disturbance produced by an infinite cylinder with the same force per unit length at each point.

### 2.3 Matching region ($a \ll \rho \ll 1$)

The velocity produced from the inner solution for $\rho \gg a$, should asymptotically match the velocity field from the outer solution for $\rho \ll 1$ as the velocity field cannot abruptly change in this matching region (i.e. $a \ll \rho \ll 1$). Matching the velocity fields from the inner and outer solutions, using equations (2.4)-(2.6) and (2.8), yields the constants in the inner solution, $\hat{B} = (U + (\omega \times r_c)) \cdot e_x + \frac{f_x}{8\pi}$ , $B = -(U + (\omega \times r_c)) \cdot e_y - \frac{f_y}{8\pi}$ and $E = -\frac{4f_z}{8\pi}$ , and leads to an integral equation for the force per unit length given by

$$\frac{f(r)}{8\pi} = \frac{\epsilon}{2}\left(I - \frac{e_z e_z}{2}\right) \cdot \left\{U + \omega \times r_c - u_\infty(r_c) - \frac{1}{8\pi}(I - 3e_z e_z) \cdot f(r) - \frac{1}{4\pi}(I + e_z e_z) \cdot \right.$$

$$f(r) \ln\left(\frac{\sqrt{1-s^2}}{\frac{a(s)}{a_0}}\right) + \frac{1}{8\pi} \int_{r_c} ds' \left[\left(\frac{I}{|r_c(s)-r_c(s')|} + \frac{(r_c(s)-r_c(s'))(r_c(s)-r_c(s'))}{|r_c(s)-r_c(s')|^3}\right) \cdot f(r') - \left(\frac{I}{|s-s'|} + \right.\right.$$

$$\left.\left.\frac{e_z e_z}{|s-s'|}\right) \cdot f(r)\right]\right\}. \quad (2.9)$$

The leading order force per unit length $f = 4\pi\epsilon\left(U + \omega \times r_c - u^\infty(r_c)\right) \cdot \left(I - \frac{e_z e_z}{2}\right)\left(1 + O(\epsilon)\right)$, suggests that a slender filament of any arbitrary cross-section experiences an $O(\epsilon)$ viscous drag equal to the viscous drag per unit length experienced by a long cylinder due to its motion relative to the local fluid velocity. The higher order terms in equation (2.9) include the additional drag due to the relative motion of the particle and the local velocity as well as a contribution that comes from the velocity disturbance created by the particle itself.

## 3. SBT solution for non-circular cross-sections

In this section the $O(\alpha\epsilon/A)$ force per unit length exerted by the filament for a slightly non-circular cross-section is derived along with the $O(\alpha\epsilon^2)$ correction to the force per unit length derived by Batchelor (1970). $\alpha$ is the perturbation parameter that quantifies the degree of non-circularity. The cross-sectional shapes that trigger these respective contributions are obtained from thought experiments. Finally, a strategy to extend our perturbation analysis to particles with $\alpha \sim O(1)$ is demonstrated towards the end of this section.

### 3.1 Regular perturbation of the inner solution

A slightly non-circular cross-section can be described by $\rho = a\big(1 + \alpha\, h(s, \theta)\big)$, where $\alpha \ll 1$, $h(s, \theta)$ is a smooth and bounded function periodic in $\theta$ with a period of $2\pi/N$, where N is any positive integer, such that $\max|h(s, \theta)| \sim O(1)$ and $|\partial h/\partial s| \sim O(1)$. The derivative $\partial h/\partial \theta$ cannot be zero, as that corresponds to a larger circular cross-section, thereby only changing the particle aspect ratio. The inner velocity field obtained in section 2 will be modified to satisfy the no slip boundary condition at the new surface, $\rho = a(1 + \alpha\, h)$. The additional velocity field in the transverse plane ($\boldsymbol{e_x} - \boldsymbol{e_y}$ plane), is described using a stream function ($\alpha\, \psi'$), such that $\psi' \sim O(1)$ for $\rho/a \sim O(1)$. Here, $\psi'$ is given by an equation of the same form as equation (2.3), but with different constants from those obtained for $\psi$. The terms in $\psi'$ corresponding to a decaying velocity field with increasing values of $\rho/a$ can be obtained by satisfying the no-slip boundary condition at $O(\alpha)$ for the velocity obtained using the stream function ($\psi + \alpha\psi'$). These constants will depend on $h$, $\tilde{\psi}$, $B$, $\hat{B}$ and $\boldsymbol{U}$ and $\boldsymbol{\omega}$ all of which are either known or obtained from the analysis done for a slender filament with a circular cross-section in section 2. The non-decaying terms in $\psi'$ used during the matching process are given by

$$\alpha\, \psi' = \alpha\left[-(C'\cos(\theta) + D'\sin(\theta))\left(\left(\frac{\rho}{a}\right)\ln\left(\frac{\rho}{a}\right)\right) + (B'\cos(\theta) + \hat{B}'\sin(\theta))\left(\left(\frac{\rho}{a}\right) - 2\ln\left(\frac{\rho}{a}\right)\left(\frac{\rho}{a}\right)\right)\dots\right],$$

<div align="right">(3.1)</div>

where $C'$ and $D'$ are constants that are obtained by satisfying the no-slip condition on the particle surface, $\rho = a(1 + \alpha h)$, while $B'$ and $\hat{B}'$ are obtained by matching with the outer solution and play the same role as $B$ and $\hat{B}$ played in the analysis for a circular cross-section. The "…" in equation (3.1) corresponds to the additional terms in the stream function necessary to satisfy the no-slip boundary condition on the particle surface that do not participate in the matching process. The corresponding terms in the fluid velocities used in the matching solution are given by

$$\alpha u'_\rho = \alpha \left[ (C' \, sin(\theta) - D' \, cos(\theta)) \left( \ln\left(\frac{\rho}{a}\right) \right) + (-B' \, sin(\theta) + \hat{B}' \, cos(\theta)) \left( 1 - 2\ln\left(\frac{\rho}{a}\right) \right) … \right],$$

(3.2)

$$\alpha u'_\theta = \alpha \left[ (C' \, cos(\theta) + D' \, sin(\theta)) \left( 1 + \ln\left(\frac{\rho}{a}\right) \right) + (B' \, cos(\theta) + \hat{B}' \, sin(\theta)) \left( 1 + 2\ln\left(\frac{\rho}{a}\right) \right) … \right].$$

(3.3)

From equation (3.2) it can be easily seen that $C'$ and $D'$ do not enter the $u_\rho$ boundary condition at $O(\alpha)$. The constants $C'$ and $D'$, which occur in the $u_\theta$ boundary condition, are given by

$$C' = \frac{1}{\pi} \int_0^{2\pi} d\theta \, \cos(\theta) \left\{ -\frac{d}{d\alpha} \left( -\frac{\partial \tilde{\psi}}{\partial \rho} \bigg|_{\rho = a(1 + \alpha \, h(s,\theta))} \right)_{\alpha=0} + 4 \, h(\theta) \left[ -\frac{f_x}{8\pi} \sin(\theta) + \frac{f_y}{8\pi} \cos(\theta) \right] \right\},$$

(3.4)

$$D' = \frac{1}{\pi} \int_0^{2\pi} d\theta \, \sin(\theta) \left\{ -\frac{d}{d\alpha} \left( -\frac{\partial \tilde{\psi}}{\partial \rho} \bigg|_{\rho = a(1 + \alpha \, h(s,\theta))} \right)_{\alpha=0} + 4 \, h(\theta) \left[ -\frac{f_x}{8\pi} \sin(\theta) + \frac{f_y}{8\pi} \cos(\theta) \right] \right\}.$$

(3.5)

The first terms in the integrands of equations (3.4) and (3.5) are of $O(1/A)$ and are related to the force per unit length driven by the gradient in the imposed fluid velocity as will be explained in section (3.2). The remainder of the O($\epsilon$) terms in the integrand of equations (3.4) and (3.5) occur due to the velocity disturbance created by the force per unit length of the unperturbed circular cross-section and are thereby driven by the motion of the particle relative to the local fluid velocity.

The function $h(s, \theta)$ is expanded as a Fourier series to understand the effect of the cross-sectional shapes that affect $C'$ and $D'$. Here, $h(s, \theta)$ is given by

$$h(s, \theta) = \sum_{m=1}^{\infty} h_m(s) \cos\big(m(\theta - \theta_{0m})\big), \tag{3.6}$$

where coefficients $h_m$, and $\theta_{0m}$, for $m = 1, 2, ...$, are constants obtained using the orthogonality of $\cos\big(m(\theta - \theta_{0m})\big)$. The effects of the Fourier modes of $h$ can be summed up to get the overall effect of the shape as the analysis is done at linear order in $\alpha$. The cross-sectional shape, $\rho = a(1 + \alpha h_m \cos(m\theta - m\theta_{0m}))$, corresponding to the $m^{th}$ Fourier mode has an $m$-fold rotational symmetry and $m$-lobes where one of the lobes makes an angle of $\theta_{0m}$ with $\boldsymbol{e_x}$. The first Fourier mode only changes the position of the cross-section without distorting the shape at linear order in $\alpha$. Thus, particles with only the first Fourier mode can be studied using the SBT formalism explained in section two, by redefining the particle centerline so that it passes through the new center of the cross-section. The other Fourier modes can affect the force per unit length in a non-trivial way. Substituting equation (3.6) in equations (3.4) and (3.5), suggests that only the second Fourier mode will affect $C'$ and $D'$ at $\mathrm{O}(\epsilon)$. The corresponding cross-sectional shape $\rho = a(1 + \alpha h_2 \cos(2\theta - 2\theta_{02}))$ is approximately an ellipse of eccentricity $\sqrt{4\alpha h_2}$ whose major axis is at an angle $\theta_{02}$ as shown in figure 1 (b). The Fourier modes of a shape that affect $C'$ and $D'$ at $O(1/A)$ depend on the specific nature of $\tilde{\psi}$ as can be seen by substituting $\rho = a(1 + \alpha h)$ in the expression for $(-\partial\tilde{\psi}/\partial\rho)$. If the imposed velocity field has terms that scale with $\rho^N$, for $N \geq 1$, then the $(N-2)^{th}$, $(N)^{th}$ and $(N+2)^{th}$ Fourier modes of $h$ will affect $C'$ and $D'$ (equation (S 1.4) of supplementary material). If $N = 1$, which corresponds to a linear imposed velocity field, then only the third Fourier mode will affect the $O(1/A)$ term in equations (3.4) and (3.5) and the corresponding three-lobed cross-sectional shape is given by $\rho = a(1 + \alpha h_3 \cos(3\theta - 3\theta_{03}))$ which is shown in figure 1 (c). If the imposed fluid velocity changes on the length scale of the particle cross-sectional size, then the contribution of the terms in $\tilde{\psi}$ which scale with $\rho^{N+1}$ to $C'$ and $D'$ will be $\mathrm{O}(1/A^N)$, which is algebraically smaller than the contribution to $C'$ and $D'$ due to a linear flow field for $N \geq 2$. Therefore, in the remainder of the paper the focus is only on a linear flow field, i.e. $\tilde{\psi}$ grows as $\rho^2$. In this case, the cross-sectional shapes shown in figures 1(b)-(d) can affect the force per unit length due to non-circularity.

Thought experiments are presented using two-dimensional Stokes flow problems to gain physical insight into why only the second and third Fourier mode perturbations to a circle change the force per unit length acting on a long cylinder. Consider a two-dimensional obstacle, whose shape is a second Fourier mode perturbation to a circle, that is placed in a uniform flow field. This body experiences a lift force, unless the imposed fluid velocity is along one of its two axes of symmetries as shown in figure 2 (a). The third and higher Fourier mode perturbations to a circle have at least a two-fold rotational symmetry in a two-dimensional (2-D) space. An $N$-fold rotational symmetry means that the shape looks the same after rotating it by any integer multiple of $2\pi/N$. These cross-sections cannot generate a lift force in a 2-D uniform flow along at least two non-colinear directions, $\boldsymbol{n_1}$ and $\boldsymbol{n_2}$ as shown in figure 2 (b), due to fore-aft symmetry. Thereby such a cross-section should experience no lift for all cross-sectional orientations by linear superposition. This is analogous to our result obtained earlier that only slender particles with cross-sections with a second Fourier mode perturbation to a circle will experience an additional force per unit length.

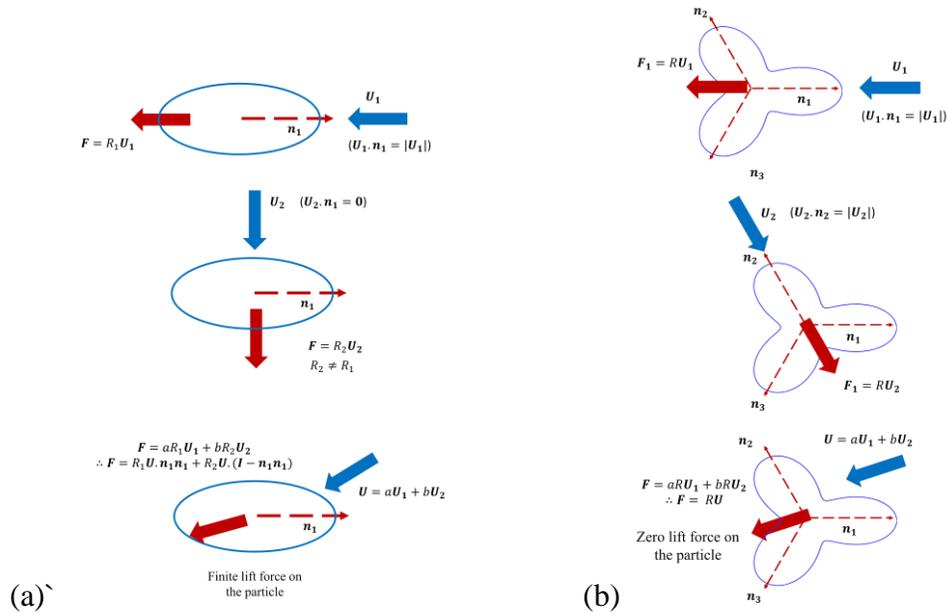

(a)`                                    (b)

Figure 2. Schematic of thought experiments to intuitively understand the importance of the second Fourier mode perturbation to a circle. (a) Finite lift on a second Fourier mode perturbation of a circle. (b) Zero lift force on a cross-section with a three lobed cross-section which is obtained from the third Fourier mode perturbation to a circle. This zero lift is true for the third and any higher Fourier mode perturbation to a circle.

To gain insight into the force per unit length driven by the gradient in the imposed fluid velocity, consider a cross-section placed in a general linear flow field and note that at $O(\alpha)$ the influence of the Fourier mode perturbations can be linearly superimposed. Only the extensional component of the flow influences the force per unit length, $\breve{\boldsymbol{f}}$. Using the linearity of Stokes equations, $\breve{\boldsymbol{f}}$ is given by

$$\breve{\boldsymbol{f}} = \xi_1 \boldsymbol{E}_\infty : \boldsymbol{n_k}\boldsymbol{n_k}\boldsymbol{n_k} + \xi_2 \boldsymbol{E}_\infty \cdot \boldsymbol{n_k}, \qquad (3.7\ a)$$

where $\boldsymbol{E}_\infty$ is the straining tensor of the linear flow field and $\boldsymbol{n_k} = \left[\cos\left(\frac{2\pi k}{n}\right), \sin\left(\frac{2\pi k}{n}\right)\right]$ with $k = 0, 1, \dots, (n-1)$ being the orientations along the lines of symmetry of the $n^{th}$ Fourier mode perturbation to the circle ($n \geq 2$). Choosing $\boldsymbol{E}_\infty = \begin{bmatrix} 0 & 1 \\ 1 & 0 \end{bmatrix}$, $\breve{\boldsymbol{f}}$ can be written as

$$\breve{\boldsymbol{f}} = \begin{bmatrix} \left(2\cos^2\left(\frac{2\pi}{n}k\right)\xi_1 + \xi_2\right)\sin\left(\frac{2\pi}{n}k\right) \\ \left(2\sin^2\left(\frac{2\pi}{n}k\right)\xi_1 + \xi_2\right)\cos\left(\frac{2\pi}{n}k\right) \end{bmatrix} = \begin{bmatrix} 0 \\ \xi_2 \end{bmatrix}_{k=0} \qquad (3.7\ b)$$

where the second equality gives the value $\breve{\boldsymbol{f}}$ at $k = 0$. $\breve{\boldsymbol{f}}$ as per equation (3.7 b) can be identical and non-zero for $\forall\, k \in \{0, 1, \dots, (n-1)\}$ only for the third Fourier mode perturbations to a circle. Identical values of $\breve{\boldsymbol{f}}$ for any other Fourier mode perturbation would require $\xi_1 = \xi_2 = 0$ implying $\breve{\boldsymbol{f}} = \boldsymbol{0}$. This implies that at linear order in the perturbation parameter, the force per unit length is only affected by the third Fourier mode perturbation to a circle. One can get a visual picture for the existence of $\breve{\boldsymbol{f}} = [0\ \xi_2]$, using our linear perturbation analysis for $u'_\theta$. For $\boldsymbol{E}_\infty = \begin{bmatrix} 0 & 1 \\ 1 & 0 \end{bmatrix}$, the perturbation to the tangential velocity $\alpha_3 u'_\theta = -\alpha_3(4\cos(2\theta)\cos(3\theta))$ is required to satisfy the no-slip condition on the particle surface. The direction of $u'_\theta$ shown in figure 3, suggests that $u'_\theta$ points along $-\boldsymbol{e_y}$ for $\theta \in \{(0, \pi/6), (\pi/4, 3\pi/4), (5\pi/6, 7\pi/6), (5\pi/4, 7\pi/4),\ (11\pi/6, 2\pi)\}$ and points along $\boldsymbol{e_y}$ for the remaining narrow portions of angular space. A net average velocity along $(-\boldsymbol{e_y})$ is required to satisfy the no-slip boundary condition which leads to the force $\check{\boldsymbol{f}} = (0, \xi_2)$ along that direction.

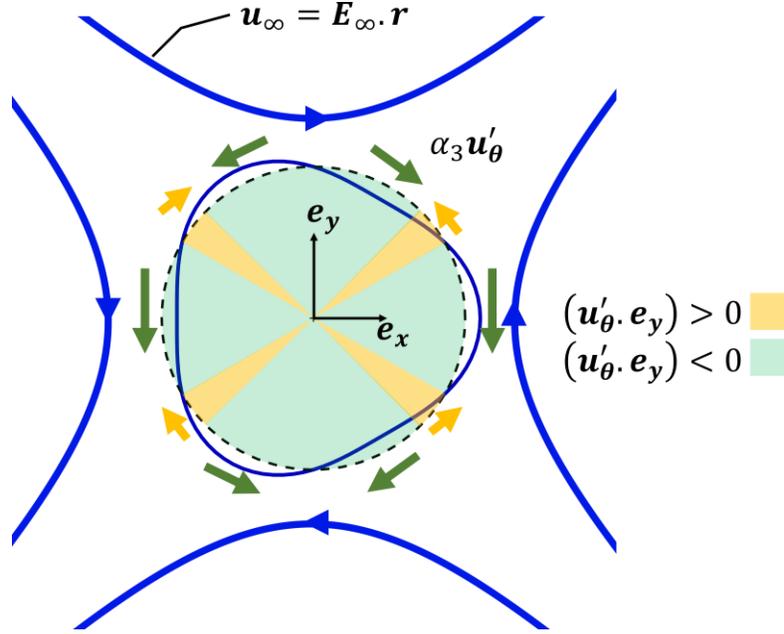

Figure 3. Direction of the perturbation to the tangential velocity on the particle surface $\alpha_3 u'_\theta = -\alpha_3(4\cos(2\theta)\cos(3\theta))$ necessary to satisfy the no-slip condition suggesting a need for a force along $\boldsymbol{e_y}$.

The additional terms in the longitudinal velocity field used during the matching process are given by

$$\alpha u'_z = \alpha e'_0 + \alpha E' \ln\left(\frac{\rho}{a}\right) + \cdots, \tag{3.8}$$

where $e'_0$ is obtained from the no-slip boundary condition and $E'$ is determined by matching the inner solution to the outer solution in a similar manner to obtaining E. Substituting $\rho = a(1 + \alpha h)$ in equation (2.8), it can be shown that there is no term of O($\alpha$) which does not depend on $\theta$, and therefore $e'_0$ is zero.

The velocity field in the inner region for a cross-section given by $\rho = a(1 + \alpha h)$, which could be obtained from equations (2.4)-(2.6), (3.2), (3.3) and (3.8), is given by

$$u_i^{inner} = u_i + \alpha u_i', \tag{3.9}$$

where $i = \{\rho, \theta, z\}$.

### 3.2 Matching the inner and outer velocity field for a slightly non-circular cross-section

Due to the linearity of Stokes equations, the additional force per unit length that arises from the perturbation of the circular cross-section must be of the form $(\alpha \boldsymbol{f}')$, where $|\boldsymbol{f}'|$ is at most $O(\epsilon^2)$. The outer velocity disturbance created by $\boldsymbol{f}'$ should also have a form similar to the velocity disturbance created by $\boldsymbol{f}$, i.e.

$$\boldsymbol{u}'(\boldsymbol{r}) \approx -\frac{1}{4\pi}(\boldsymbol{I} + \boldsymbol{e}_z\boldsymbol{e}_z) \cdot \boldsymbol{f}'(\boldsymbol{r}) \left[ \ln\left(\frac{\rho}{2}\right) - \ln\left(\sqrt{1-s^2}\right) \right] - \frac{1}{4\pi}\boldsymbol{f}' \cdot \boldsymbol{e}_z\boldsymbol{e}_z + \frac{1}{4\pi}\boldsymbol{f}' \cdot \boldsymbol{e}_\rho\boldsymbol{e}_\rho +$$

$$\frac{1}{8\pi}\int_{r_c} ds' \left[ \left( \frac{\boldsymbol{I}}{|\boldsymbol{r}_c(s)-\boldsymbol{r}_c(s')|} + \frac{\big(\boldsymbol{r}_c(s)-\boldsymbol{r}_c(s')\big)\big(\boldsymbol{r}_c(s)-\boldsymbol{r}_c(s')\big)}{|\boldsymbol{r}_c(s)-\boldsymbol{r}_c(s')|^3} \right) \cdot \boldsymbol{f}'(\boldsymbol{r}') - \left( \frac{\boldsymbol{I}}{|s-s'|} + \frac{\boldsymbol{e}_z\boldsymbol{e}_z}{|s-s'|} \right) \cdot \boldsymbol{f}'(\boldsymbol{r}) \right]. \tag{3.10}$$

The complete velocity field, $\boldsymbol{u} + \alpha\boldsymbol{u}'$, is obtained by combining equations (2.8) and (3.10). On matching the velocity fields from the inner and outer region, $\boldsymbol{f}'$ is given by the integral equation:

$$\boldsymbol{f}'(\boldsymbol{r}) = \frac{\epsilon}{2}\left(\boldsymbol{I} - \frac{\boldsymbol{e}_z\boldsymbol{e}_z}{2}\right) \cdot \left( \frac{C'}{2}\boldsymbol{e}_y - \frac{D'}{2}\boldsymbol{e}_x - \frac{1}{8\pi}(\boldsymbol{I} - 3\boldsymbol{e}_z\boldsymbol{e}_z) \cdot \boldsymbol{f}'(\boldsymbol{r}) - \frac{1}{4\pi}(\boldsymbol{I} + \boldsymbol{e}_z\boldsymbol{e}_z) \cdot \right.$$

$$\boldsymbol{f}'(\boldsymbol{r}) \ln\left( \frac{\sqrt{1-s^2}}{\frac{a(s)}{a_0}} \right) + \frac{1}{8\pi}\int_{r_c} ds' \left[ \left( \frac{\boldsymbol{I}}{|\boldsymbol{r}_c(s)-\boldsymbol{r}_c(s')|} + \frac{\big(\boldsymbol{r}_c(s)-\boldsymbol{r}_c(s')\big)\big(\boldsymbol{r}_c(s)-\boldsymbol{r}_c(s')\big)}{|\boldsymbol{r}_c(s)-\boldsymbol{r}_c(s')|^3} \right) \cdot \boldsymbol{f}'(\boldsymbol{r}') - \left( \frac{\boldsymbol{I}}{|s-s'|} + \right.\right.$$

$$\left.\left. \frac{\boldsymbol{e}_z\boldsymbol{e}_z}{|s-s'|} \right) \cdot \boldsymbol{f}'(\boldsymbol{r}) \right] \right). \tag{3.11}$$

Here, $\boldsymbol{f}'$ can be iteratively obtained with errors of $O(\epsilon^{N+1})$ or errors of $O(\epsilon^N/A)$ if $h_2 = 0$, where $N \geq 2$. $\boldsymbol{f}'$ does not have a component along the longitudinal direction at linear order in $\alpha$. This can be understood on matching the $O(\alpha)$ terms in the inner and outer velocity fields in $u_z'$, which results in $E'$ being identically zero.

The integral equation for the net force per unit length, $\boldsymbol{f_{net}} = \boldsymbol{f} + \alpha \boldsymbol{f}'$ is given by

$$\frac{(f_{net})}{8\pi} = \frac{\epsilon}{2}\left(\boldsymbol{I} - \frac{\boldsymbol{e_z e_z}}{2}\right) \cdot \left\{\boldsymbol{U} + \boldsymbol{\omega} \times \boldsymbol{r_c} - \boldsymbol{u_\infty}(\boldsymbol{r_c}) + \alpha\left(\frac{C'}{2}\boldsymbol{e_y} - \frac{D'}{2}\boldsymbol{e_x}\right) - \frac{1}{8\pi}(\boldsymbol{I} - 3\boldsymbol{e_z e_z}) \cdot\right.$$

$$(f_{net}) - \frac{1}{4\pi}(\boldsymbol{I} + \boldsymbol{e_z e_z}) \cdot (f_{net}) \ln\left(\frac{\sqrt{1-s^2}}{\frac{a(s)}{a_0}}\right) + \frac{1}{8\pi}\int_{r_c} ds' \left[\left(\frac{\boldsymbol{I}}{|\boldsymbol{r_c}(s) - \boldsymbol{r_c}(s')|} + \right.\right.$$

$$\left.\frac{\left(\boldsymbol{r_c}(s) - \boldsymbol{r_c}(s')\right)\left(\boldsymbol{r_c}(s) - \boldsymbol{r_c}(s')\right)}{|\boldsymbol{r_c}(s) - \boldsymbol{r_c}(s')|^3}\right) \cdot \left(\boldsymbol{f_{net}}(\boldsymbol{r'})\right) - \left(\frac{\boldsymbol{I}}{|s - s'|} + \frac{\boldsymbol{e_z e_z}}{|s - s'|}\right) \cdot \left(\boldsymbol{f_{net}}(\boldsymbol{r})\right)\left]\right\},$$  (3.12)

where $C'$ and $D'$ are obtained from equations (3.2) and (3.3) respectively and have contributions of order $\epsilon$ and/or $1/A$ depending on the shape of the cross-section. The governing integral equation can be completely solved to get the value of the force per unit length correct to O($\epsilon^N$) + O($\alpha\epsilon^{N+1}$) + O($\alpha\epsilon^N/A$), where N is an integer greater that unity. Solving for the force per unit length to O($\epsilon$) + O($\alpha\epsilon^2$) + O($\alpha\epsilon/A$), equation (3.12) simplifies to

$$\boldsymbol{f} + \alpha \, \boldsymbol{f}' = 4\pi\epsilon(\boldsymbol{U} + \boldsymbol{\omega} \times \boldsymbol{r_c} - \boldsymbol{u_\infty}(\boldsymbol{r_c})) \cdot \left(\boldsymbol{I} - \frac{\boldsymbol{e_z e_z}}{2}\right) + 2\pi\epsilon\alpha\left(-D'\boldsymbol{e_x} + C'\boldsymbol{e_y}\right).$$  (3.13)

Equation (3.13) suggests that the force per unit length experienced by a slender filament due to non-circularity of the cross-section is affected at O($\alpha\epsilon^2$) due to the velocity disturbance of the unperturbed circular cross-section and at O($\alpha\epsilon/A$) due to the gradient in the imposed fluid velocity.

### 3.3 Extending the analysis to a general cross-sectional shape ($\alpha \sim O(1)$)

Here, a numerical calculation to determine the velocity disturbance by any cross-section is elucidated which can be used as the matching solution for SBT. This calculation involves the solution to the flow past an obstacle with the same shape as the particle cross-section in a two-dimensional domain with a size, $\rho_\infty$, that is much larger than the cross-sectional size ($\rho_\infty \gg a$) as shown in figure 4. In this subsection, the fluid viscosity, a measure of the undisturbed fluid velocity far away from the obstacle and a length that is of the order of the size of the obstacle are used to non-dimensionalize any quantity of interest, such as the force per unit length.

The apparent hydrodynamic center of resistance (AHCOR[1]) of the cross-section is chosen as the center of the computational domain to avoid a solid body rotation at large separations from the particle. The AHCOR is defined as the point about which zero torque is acting on a two-dimensional obstacle translating in a concentric circular domain of size much larger than the obstacle with the outer boundary having zero velocity.

Consider a stationary obstacle experiencing a force per unit length $\boldsymbol{f}$ placed in a fluid with an imposed velocity of $\boldsymbol{u}_\infty \cdot (\boldsymbol{I} - \boldsymbol{e}_z \boldsymbol{e}_z)$. The velocity field at the outer boundary is obtained from the asymptotic form predicted by SBT according to equations (2.8) and (3.11). This asymptotic form of the imposed fluid velocity in the region $\rho \gg a$ is given by

$$\boldsymbol{u} = \boldsymbol{u}_\infty + \frac{2f}{8\pi} \cdot \left[ -\ln\left(\frac{\rho}{a}\right)(\boldsymbol{I} + \boldsymbol{e}_z \boldsymbol{e}_z) + \left(\boldsymbol{e}_\rho \boldsymbol{e}_\rho - 0.5(\boldsymbol{I} - \boldsymbol{e}_z \boldsymbol{e}_z)\right) + \boldsymbol{K} \right] + \boldsymbol{L}, \qquad (3.14)$$

where $\boldsymbol{K}$ is a second-order tensor that depends only on the geometry of the particle cross-section and $\boldsymbol{L}$ is a vector that arises due to the gradient of the imposed fluid velocity and therefore depends both on the geometry of the particle and $\boldsymbol{u}_\infty$. $\boldsymbol{K}$ is symmetric (Batchelor 1970), such that $K_{iz} = K_{zi} = 0$ for $i = \{x, y\}$, due to the decoupling of the longitudinal and the transverse flow fields around a long slender body. The corresponding pressure for this velocity field in the region $\rho \gg a$ is given by

$$p = \frac{4}{8\pi\rho} \boldsymbol{f} \cdot \boldsymbol{e}_\rho. \qquad (3.15)$$

In equation (3.14), $\boldsymbol{K}$ and $\boldsymbol{L}$ are to be determined as part of the solution and therefore the velocity field at the outer boundary for a given $\boldsymbol{f}$ and $\boldsymbol{u}_\infty$ cannot be specified a priori. Instead, the force per unit area $\boldsymbol{t}$ acting on the outer boundary $\rho = \rho_\infty$ is specified. $\boldsymbol{t}$ is independent of $a$, $\boldsymbol{K}$ and $\boldsymbol{L}$, and given by

$$\boldsymbol{t} = \boldsymbol{t}_\infty - \frac{1}{\pi \rho_\infty} \boldsymbol{f} \cdot (\boldsymbol{e_\rho} \boldsymbol{e_\rho}), \qquad (3.16)$$

where $\boldsymbol{t}_\infty$ is the force per unit area that would act on the outer boundary due to $\boldsymbol{u}_\infty \cdot (\boldsymbol{I} - \boldsymbol{e_z} \boldsymbol{e_z})$ in the absence of the obstacle.

---

[1] AHCOR is similar to the hydrodynamic center of resistance for three dimensional particles, which is defined as the point about which the torque acting on a body translating in a quiescent fluid is zero (Kim and Karrila, 1991). In a two-dimensional Stokes flow the velocity disturbance due to an obstacle grows logarithmically with the domain size and therefore a hydrodynamic center of resistance based on the above definition cannot be defined in the same manner.

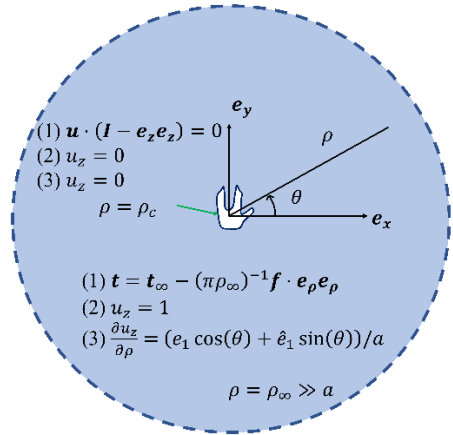

Figure 4. Schematic of the two-dimensional domain to obtain $\boldsymbol{K}$ and $\boldsymbol{L}$, along with the boundary conditions for (1) the 2-D Stokes flow equations in the transverse plane and the Laplace equation for the longitudinal velocity field $\nabla^2 u_z = 0$ for computing (2) $K_{zz}$ and (3) $L_z$.

$a$ and $K_{ij}$ for $i, j = \{x, y\}$ are obtained by solving the 2-D Stokes equations with no-slip on the obstacle surface and $\boldsymbol{t} = -(1/\pi\rho_\infty) \boldsymbol{f} \cdot (\boldsymbol{e_\rho} \boldsymbol{e_\rho})$ on the outer boundary and matching the numerical velocity and $\boldsymbol{u} \cdot (\boldsymbol{I} - \boldsymbol{e_z} \boldsymbol{e_z})$ with $\boldsymbol{L} = 0$ at the outer boundary. $L_i$ for $i = \{1,2\}$ is calculated by equating $\boldsymbol{u} = (\boldsymbol{u}_\infty + \boldsymbol{L}) \cdot (\boldsymbol{I} - \boldsymbol{e_z} \boldsymbol{e_z})$ to the velocity field at the outer boundary obtained from the solution of the 2-D Stokes equations with no slip on the obstacle surface and $\boldsymbol{t} = \boldsymbol{t}_\infty$ at $\rho = \rho_\infty$. $u_z$ at the outer boundary obtained from the solution of $\nabla^2 u_z = 0$, with no-slip on the obstacle and a constant value at $\rho = \rho_\infty$, is equated to $u_z$ from equation (3.14) to obtain $K_{zz}$. Similarly, $u_z$ at the outer boundary obtained from the solution of $\nabla^2 u_z = 0$, with no slip on the obstacle and $\partial u_z / \partial \rho$ corresponding to a linear vector field $\boldsymbol{u}_\infty \cdot \boldsymbol{e_z}$ at $\rho = \rho_\infty$, is equated to $u_z$ from equation (3.14) to obtain $L_z$. For arbitrary cross-sections the 2-D Stokes flow equations and the 2-D Laplace's equation can be solved using a finite element solver such as COMSOL or a two-dimensional boundary element method, by choosing a $\rho_\infty$ which is sufficiently large such that the values of $a, K$ and $L$ do not change on further increasing $\rho_\infty$. The final integral equation for the force per unit length exerted by a slender filament with an arbitrary cross-section is given by

$$\frac{f(r)}{8\pi} = \frac{\epsilon}{2}\left(I - \frac{e_z e_z}{2}\right) \cdot \left\{ U + \omega \times r_c - u_\infty(r_c) + \frac{1}{4\pi}f(r) \cdot K + L - \frac{1}{8\pi}(I - 3e_z e_z) \cdot f(r) - \frac{1}{4\pi}(I + e_z e_z) \cdot f(r)\ln\left(\frac{\sqrt{1-s^2}}{\frac{a(s)}{a_0}}\right) + \frac{1}{8\pi}\int_{r_c} ds' \left[ \left( \frac{I}{|r_c(s) - r_c(s')|} + \frac{\left(r_c(s) - r_c(s')\right)\left(r_c(s) - r_c(s')\right)}{|r_c(s) - r_c(s')|^3} \right) \cdot f(r') - \left( \frac{I}{|s-s'|} + \frac{e_z e_z}{|s-s'|} \right) \cdot f(r) \right] \right\}.$$

(3.17)

Note that $K$ is related to the tensor $K_B$ and $\ln(k_B)$, which Batchelor (1970) mentions in equations (5.5) and (6.1) of his paper respectively, by the relation $K_B + 2\ln(k_B) e_z e_z = K + (I + e_z e_z)\ln(a/R_B)$, where $R_B$ is such that $2\pi R_B$ is the perimeter of the cross-section. Batchelor (1970) shows that $K_B$ is a symmetric tensor, which implies that $K$ is also symmetric. $K_B$ has three unknowns such that $K_B \cdot e_z = 0$. These unknowns are written in terms of $a$ and $K_{ij}$, where $i, j = \{x, y\}$, such that $K_{xx} + K_{yy} = 0$. Also, $K_{zz} = 2\ln(k_B) - 2\ln(a/R_B)$ captures the effect of the cross-sectional shape on the force and velocity field in the longitudinal direction. The influence of the cross-section on the velocity disturbance in the transverse plane can be decomposed into an isotropic component $\ln(a) I$ and a traceless component $K - K_{zz} e_z e_z$. This length scale "$a$" which arises as part of the hydrodynamic calculation is therefore used instead of a purely geometric length scale $R_B$.

The second-order tensor, $K$ can be represented in terms of $\bar{\alpha}_2$, $\theta_{02}$ and $K_{zz}$ and is given by

$$K = \begin{bmatrix} -0.5\bar{\alpha}_2\cos(2\theta_{02}) & -0.5\bar{\alpha}_2\sin(2\theta_{02}) & 0 \\ -0.5\bar{\alpha}_2\sin(2\theta_{02}) & 0.5\bar{\alpha}_2\cos(2\theta_{02}) & 0 \\ 0 & 0 & K_{zz} \end{bmatrix}.$$

(3.18 a)

$(\bar{\alpha}_2, \theta_{02})$, similar to $(\alpha_2, \theta_{02})$ of the regular perturbation theory, give the contribution to the second Fourier mode of the cross-sectional geometry. $K_{zz}$, which is zero to leading order in the perturbation analysis, can have a non-zero value for a general cross-sectional shape. This is because $u_z$ can be affected by the details of the cross-section at $O(\alpha^2)$ (See supplementary material S.2 for details). The vector $\boldsymbol{L}$ can represented in terms of $\bar{\alpha}_3$, $\theta_{03}$, $L_z$ and the imposed fluid velocity and is given by

$$\boldsymbol{L} = \begin{bmatrix} 2\bar{\alpha}_3(\hat{a}_2\cos(3\theta_{03}) - a_2\sin(3\theta_{03})) \\ 2\bar{\alpha}_3(a_2\cos(3\theta_{03}) + \hat{a}_2\sin(3\theta_{03})) \\ L_{zx}e_1 + L_{zy}\hat{e}_1 \end{bmatrix}, \qquad (3.18\ b)$$

where the imposed velocity field $(\boldsymbol{u}_\infty(\boldsymbol{r}) - \boldsymbol{u}_\infty(\boldsymbol{r_c})) \cdot (\boldsymbol{I} - \boldsymbol{e_z}\boldsymbol{e_z})$ is specified in terms of a stream function $\psi_\infty$ given by

$$\frac{\psi_\infty}{a} = \tilde{a}_0\left(\frac{\rho}{a}\right)^2 + (a_2\cos(2\theta) + \hat{a}_2\sin(2\theta))\left(\frac{\rho}{a}\right)^2, \qquad (3.18\ c)$$

and $(\boldsymbol{u}_\infty(\boldsymbol{r}) - \boldsymbol{u}_\infty(\boldsymbol{r_c})) \cdot \boldsymbol{e_z}\boldsymbol{e_z}$ is given by

$$u_{\infty,z} = (e_1\cos(\theta) + \hat{e}_1\sin(\theta))\frac{\rho}{a}. \qquad (3.18\ d)$$

$(\bar{\alpha}_3, \theta_{03})$, similar to $(\alpha_3, \theta_{03})$ of the regular perturbation theory, give the contribution to the third Fourier mode perturbation. The longitudinal component $\boldsymbol{L} \cdot \boldsymbol{e_z}$ is zero in the linear perturbation analysis but it is non-zero for a general cross-sectional geometry. The longitudinal component of $-\boldsymbol{L}$ is the longitudinal velocity at which a particle must translate to avoid a longitudinal force per unit length when it is subjected to a simple shear flow with the stagnation streamline coinciding with the AHCOR. For the geometries studied in this paper, $\boldsymbol{L} \cdot \boldsymbol{e_z}$ was found to be numerically small compared to the components of $\boldsymbol{L}$ in the transverse plane.

## 4. Resistance to motion of a triaxial ellipsoid

In this section our theory is utilized to obtain the Stokes hydrodynamic resistance tensor of triaxial ellipsoids of semi-axis lengths $l_1, l_2$ and $l_3$, such that $l_3 \gg l_1 > l_2$ (3 is the longitudinal direction, 1 is along the long axis of the elliptical cross-section and directions, [1,2,3] form a right-handed Cartesian coordinate system). By symmetry of the shape, the force, $\boldsymbol{F}$ and the torque, $\boldsymbol{T}$ acting on the ellipsoid are given as $\boldsymbol{F} = \boldsymbol{R}^{FU} \cdot \boldsymbol{U}$, $\boldsymbol{T} = \boldsymbol{R}^{L\omega} \cdot \boldsymbol{\omega}$, where $\boldsymbol{R}^{FU}$ and $\boldsymbol{R}^{L\omega}$ are $3 \times 3$ diagonal matrices that depend only on the particle geometry. The values of $\boldsymbol{R}^{FU}$ and $\boldsymbol{R}^{L\omega}$ for $l_3/l_1 \gg 1$ and $(l_1 - l_2)/l_1 \ll 1$ using the perturbation analysis in section (3.2) are given by

$$\boldsymbol{R}^{FU} = (8\pi) \frac{\epsilon\,(2l_3)}{2+\epsilon} \begin{bmatrix} 1 - \frac{\alpha_2 \epsilon}{2+\epsilon} & 0 & 0 \\ 0 & 1 + \frac{\alpha_2 \epsilon}{2+\epsilon} & 0 \\ 0 & 0 & \frac{1}{2}\frac{2+\epsilon}{2-\epsilon} \end{bmatrix}, \tag{4.1}$$

$$\boldsymbol{R}^{L\omega} = (8\pi) \frac{2l_3^3}{3} \frac{\epsilon}{2-\epsilon} \begin{bmatrix} 1 + \frac{\alpha_2 \epsilon}{2-\epsilon} & 0 & 0 \\ 0 & 1 - \frac{\alpha_2 \epsilon}{2-\epsilon} & 0 \\ 0 & 0 & \frac{(2-\epsilon)}{\epsilon}\left(\frac{a}{l_3}\right)^2 \end{bmatrix}, \tag{4.2}$$

where $\epsilon = 1/\ln(2l_3/a)$. $a$ and $\alpha_2$ are given by $a = 0.5(l_1 + l_2)$ and $\alpha_2 = (l_1 - l_2)/(2a)$ for $(l_1 - l_2) \ll l_1$. The results of equations (4.1) and (4.2) match exactly with the values obtained by Batchelor (1970) (equations (8.7), (8.8) and (8.10) of his paper).

For cross-sections with $(l_1 - l_2) \sim O(l_1)$, $\boldsymbol{R}^{FU}$ and $\boldsymbol{R}^{L\omega}$ are determined using the numerical procedure in section (3.3) and the results retain great accuracy even for cross-sections with extreme aspect ratio. COMSOL, a finite element solver, was used to perform the 2-D Stokes flow calculation. $\boldsymbol{K}$ was estimated with an uncertainty of below 0.1 % when the size of the outer boundary ($\rho_\infty$) was at least 30 times the cross-sectional dimension. Figure 5 shows the deviation of $\boldsymbol{R}^{FU}$ and $\boldsymbol{R}^{L\omega}$ predicted by our numerical procedure from the exact result for an ellipsoid given by Lamb (1932) for a high-aspect ratio elliptical cross-section with $l_1/l_2 = 10$. The deviation of the SBT for a circular cross-section from the exact result is also presented for comparison. Our SBT predicts $\boldsymbol{R^{FU}}$ and $\boldsymbol{R^{L\omega}}$ better than the SBT results for a circular cross-section and has errors less than around 1% for $l_3/l_1 \gtrsim 10$. The high level of accuracy shows the applicability of our methodology to accurately predict the resistance to motion of slender bodies with arbitrary cross-section.

The value of $\boldsymbol{R^{L\omega}} : \boldsymbol{e_3}\boldsymbol{e_3}$ needs special attention for a straight slender body because the force per unit length cannot generate a torque about its longitudinal axis. To obtain the effect of the cross-sectional geometry on $\boldsymbol{R^{L\omega}} : \boldsymbol{e_3}\boldsymbol{e_3}$, a two-dimensional Stokes flow problem is solved to find the torque per unit length $\boldsymbol{g} \propto a^2(s)$ acting on an ellipse with sides $l_1$ and $l_2$ which is rotating with a unit angular velocity parallel to $\boldsymbol{e_3}$ with the velocity on the outer boundary set to zero (see figure 3). The torque per unit length is integrated by accounting for the variation of the cross-sectional size to attain the total torque on the ellipsoid, and thereby obtain $\boldsymbol{R^{L\omega}} : \boldsymbol{e_3}\boldsymbol{e_3}$ as depicted in figure 5 (f).

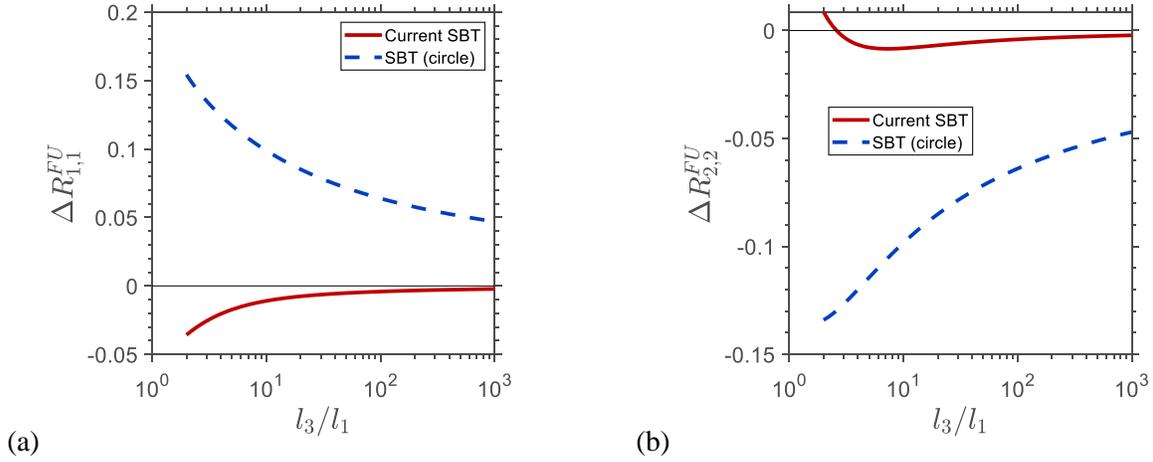

(a)                                          (b)

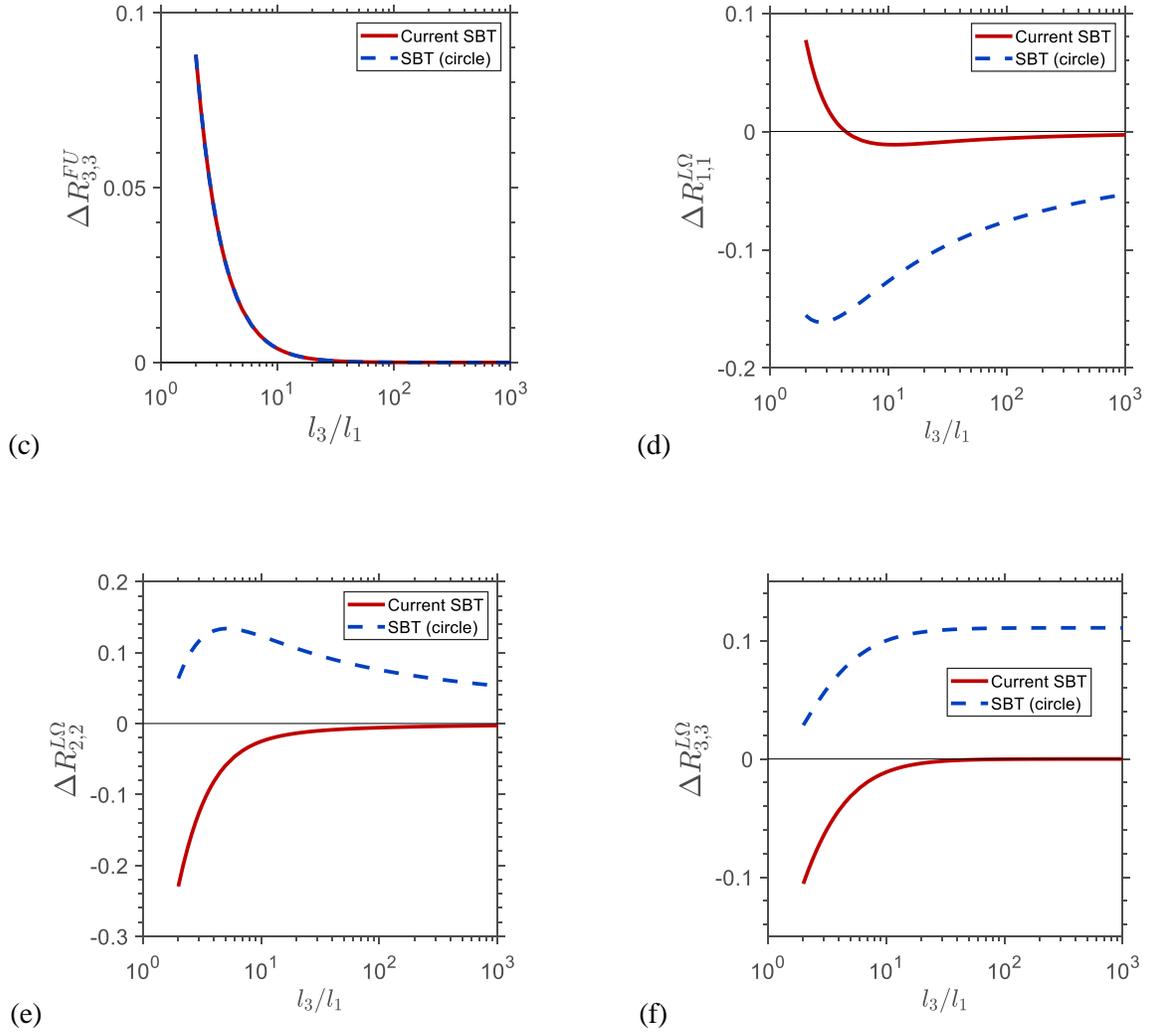

Figure 5. Variation of resistances, $\boldsymbol{R^{FU}}$ and $\boldsymbol{R^{L\omega}}$, with $l_3/l_1$ for an ellipsoid with $l_1/l_2 = 10$. (a) − (f) Comparison of $\Delta R_{ij}^k = \left(R_{ij}^k - \left(R_{ij}^k\right)_{exact}\right)/\left(R_{ij}^k\right)_{exact}$, the deviation of different components of $\boldsymbol{R^{FU}}$ and $\boldsymbol{R^{L\omega}}$ predicted using the current SBT as well as the SBT for a circuler cross-section from the exact result of Lamb (1932) ($\left(R_{ij}^k\right)_{exact}$). Here $i, j = \{1, 2, 3\}$ and $k = \{\boldsymbol{FU}, \boldsymbol{L\omega}\}$.

## 5. Translation of a straight slender body in a simple shear flow (SSF)

An axisymmetric straight particle rotates periodically in one of the Jeffery orbits depending on its initial orientation and has zero cross-stream drift relative to the fluid velocity at its center of mass. A straight particle with a three-lobed cross-section shown in figure 6 (a) rotates like a spheroid but translates quasi-periodically across streamlines with an $O(\alpha/A)$ velocity. A straight particle with a combination of an elliptic (or two-lobed) and a three-lobed cross-section can rotate chaotically and translate diffusively. The calculations in this section can be used to extract the motion of straight fibers in viscous fluids which is important in the manufacturing process of fiber-reinforced composite materials or paper products. The half-length of the particle, the shear rate and the fluid viscosity are used to non-dimensionalize variables in this section.

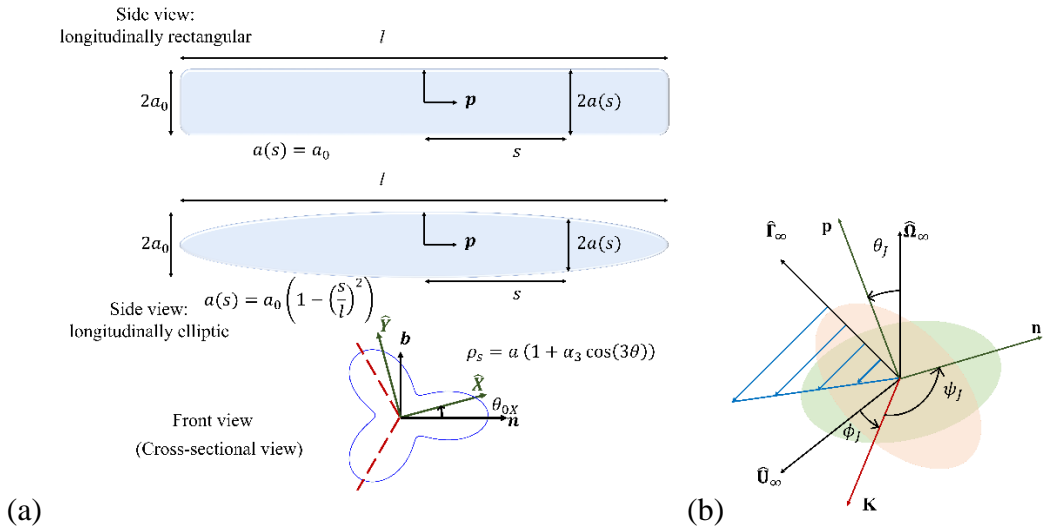

(a)                                         (b)

Figure 6. Schematic of the particle shape and the coordinate system used in the calculation. (a) Schematic of the slender particle that is longitudinally rectangular or elliptic. The cross-section of the particle is given by $\rho_s = a(1 + \alpha_3 \cos(3\theta))$, where $\theta$ is measured from $\boldsymbol{n}$. $\boldsymbol{n}$ is along one of the lines of symmetry of the cross-section, $\hat{\boldsymbol{Y}}$ is a unit vector in the cross-sectional plane chosen such that $\hat{\boldsymbol{Y}} \cdot \hat{\boldsymbol{U}}_\infty = 0$ and $\hat{\boldsymbol{X}} = \hat{\boldsymbol{Y}} \times \boldsymbol{p}$. (b) The fixed reference frame is defined along the flow ($\hat{\boldsymbol{U}}_\infty$), vorticity ($\hat{\boldsymbol{\Omega}}_\infty$) and the gradient ($\hat{\boldsymbol{\Gamma}}_\infty$) direction of the SSF. The longitudinal direction of the the particle is along $\boldsymbol{p}$ and a reference vector in the transverse plane is $\boldsymbol{n}$. $\boldsymbol{p}, \hat{\boldsymbol{\Omega}}_\infty, \boldsymbol{K} = \boldsymbol{p} \times \hat{\boldsymbol{\Omega}}_\infty, \boldsymbol{n}$ and $\hat{\boldsymbol{U}}_\infty$ are used to define the Euler angles $\theta_J, \phi_J, \psi_J$.

## 5.1. Problem formulation and SBT solution

The cross-section of the slender particle studied here is a small perturbation to a circle given by $\rho_s = a(1 + \alpha_3 \cos(3\theta))$, where $\alpha_3 \ll 1$ is the perturbation parameter and $\theta$ is measured relative to a vector $\boldsymbol{n}$ that is along one of the lines of symmetry of the cross-section. The second Fourier mode perturbation to a circle does not affect $C'$ and $D'$ for a torque-free particle in a SSF because the force per unit length it produces satisfies $\boldsymbol{f} \cdot (\boldsymbol{I} - \boldsymbol{pp}) = 0$. The second Fourier mode perturbation to a circle affects the torque per unit length $\boldsymbol{g}$ thereby influencing the rotational dynamics of a straight particle, which is discussed in section 5.3.

The size of the cross-section, "$a$", either varies with the longitudinal position, $s$, as $\frac{a(s)}{a_0} = (1 - s^2)^{0.5}$ for a cross-section that is longitudinally elliptic or is a constant $a = a_0$ for a cylinder. The force per unit length obtained from equation (3.17), is used to obtain the linear ($\boldsymbol{U}$) and angular ($\boldsymbol{\omega}$) velocity of the particle by applying the force-free ($\int \boldsymbol{f}_{net} ds = 0$) and torque-free ($\int (s\boldsymbol{p} \times \boldsymbol{f}_{net} + \boldsymbol{g}) ds = 0$) condition on the particle respectively. Here, $\boldsymbol{g}$ is calculated by computing the stresses from a transverse velocity field obtained from the stream function $\tilde{\psi}$ and the velocity field $\tilde{u}_z$.

The angular velocity $\boldsymbol{\omega}$ is not affected by a three-lobed perturbation of a circle at linear order in $\alpha_3$ and thus, this particle rotates periodically, like a spheroid shown by Jeffery (1922). This holds true for cylinders with blunt ends and a three-lobed cross-section. The ends of a blunt cylinder significantly influence $\boldsymbol{\omega}$ when the particle is near the flow-vorticity plane (Cox 1971), which can be computed using the force generated at the ends of the particle in the transverse direction, $\boldsymbol{F}_{end}$. Using linearity of Stokes flow, and the symmetry of the third and higher Fourier mode perturbations, $\boldsymbol{g} \cdot (\boldsymbol{I} - \boldsymbol{pp})$ and $\boldsymbol{p} \times \boldsymbol{F}_{end}$ can be shown to be proportional to $\boldsymbol{E} \cdot \boldsymbol{p} \times \boldsymbol{p}$ which is proportional to $\boldsymbol{\omega}$ due to the straining component of a SSF for an axisymmetric particle (See supplementary material S2). Therefore, straight particles with $\alpha_2 = 0$, in addition to circular cylinders shown by Cox (1971), rotate similar to an $O(A/\sqrt{\ln(A)})$ aspect ratio spheroid. The exact relationship can be obtained from experiments or a numerical calculation. For a torque-free straight particle with a 3-lobed cross-section, $\boldsymbol{\omega}$ is given by

$$\boldsymbol{\omega} = \boldsymbol{\omega}_\infty + \lambda_J \boldsymbol{p} \times (\boldsymbol{E} \cdot \boldsymbol{p}), \tag{5.1}$$

where $\lambda_J$ is the rotation parameter of the particle that depends only on its geometry. The rotating parameter $\lambda_J = 1 - \frac{2}{A^2}$ for a slender particle that is longitudinally elliptic (Jeffery 1922, Cox 1971) and $\lambda_J = 1 - 0.65 \frac{\ln(A)}{A^2}$ for a cylinder, where the prefactor of 0.65 was obtained by fitting the asymptotic form of Cox (1971) to the experimental data of Anczurowskei and Mason (1968).

Unlike a spheroid or a circular cylinder, a slender particle with a three-lobed cross-section drifts across streamlines due to the $O(\alpha_3/A)$ force per unit length. The drift velocity of the particle is confined to the plane normal to $\boldsymbol{p}$, as $\boldsymbol{f_{net}} \cdot \boldsymbol{p} = 0$. Thus, the drift velocity of this particle takes the form $\boldsymbol{U_p} = U_x \widehat{\boldsymbol{X}} + U_y \widehat{\boldsymbol{Y}}$, where $U_x$ and $U_y$ are the components of the drift velocity along $\widehat{\boldsymbol{X}}$ and $\widehat{\boldsymbol{Y}}$ respectively and $(\widehat{\boldsymbol{X}}, \widehat{\boldsymbol{Y}}, \boldsymbol{p})$ form an orthogonal pair such that $\widehat{\boldsymbol{Y}} \cdot \widehat{\boldsymbol{U}}_\infty = 0$. $U_x$ and $U_y$ are given by

$$U_x = \alpha_3 a_0 U_0 \big[ \boldsymbol{E}_\infty : \widehat{\boldsymbol{X}}\widehat{\boldsymbol{Y}} \sin(3\theta_{0X}) - 0.5 \, \boldsymbol{E}_\infty : \widehat{\boldsymbol{X}}\widehat{\boldsymbol{X}} \cos(3\theta_{0X}) \big], \tag{5.2 a}$$

$$U_y = \alpha_3 a_0 U_0 \big[ \boldsymbol{E}_\infty : \widehat{\boldsymbol{X}}\widehat{\boldsymbol{Y}} \cos(3\theta_{0X}) + 0.5 \, \boldsymbol{E}_\infty : \widehat{\boldsymbol{X}}\widehat{\boldsymbol{X}} \sin(3\theta_{0X}) \big], \tag{5.2 b}$$

where $U_0 = 1$ for a cylinder and $U_0 = \pi/4$ for a spheroid, $\theta_{0X}$ is the angle made by $\widehat{\boldsymbol{X}}$ with $\boldsymbol{n}$ and $\boldsymbol{E}_\infty = \frac{1}{2}(\boldsymbol{\nabla}\boldsymbol{u}^\infty + (\boldsymbol{\nabla}\boldsymbol{u}^\infty)^T)$ is the straining tensor. $\boldsymbol{U}$ for the longitudinally elliptic particle differs from $\boldsymbol{U}$ of a cylinder by a factor of $\pi/4$ due to the difference in the integral of $\int_{-1}^{1} ds \, a(s)$ for the two cases. Only the results for a straight cylinder are presented in the following section, since both $\boldsymbol{U}$ and $\boldsymbol{\omega}$ are qualitatively similar for a longitudinally elliptic particle.

### 5.2. Quasi-periodic translation of particles

The motion of the particle shown in figure 6(a) is calculated by tracking its center of mass position and orientation. The orientation is given in terms of the Euler angles, $(\theta_J, \psi_J, \phi_J)$, shown in figure 6 (b) and defined using the longitudinal direction of the particle $\boldsymbol{p} = (\sin(\phi_J)\sin(\theta_J), \cos(\theta_J), \cos(\phi_J)\sin(\theta_J))$, a vector $\boldsymbol{K} = (\cos(\phi_J), 0, -\sin(\phi_J))$ that is normal to both $\boldsymbol{p}$ and $\widehat{\boldsymbol{\Omega}}_\infty$ and $\cos(\psi_J) = \boldsymbol{n} \cdot \boldsymbol{K}$, where $\boldsymbol{n}$ is a vector along one of the lines of symmetry of the cross-section as illustrated in figure 6 (a). Here, $(\boldsymbol{n}, \boldsymbol{b}, \boldsymbol{p})$ and $(\widehat{\boldsymbol{X}}, \widehat{\boldsymbol{Y}}, \boldsymbol{p})$, such that $\widehat{\boldsymbol{Y}} \cdot \widehat{\boldsymbol{U}}_\infty = 0$, form an orthogonal set. The angle $\theta_{0X}$ in equation (5.4) equals $\mathrm{asin}(\widehat{\boldsymbol{Y}} \cdot \boldsymbol{n})$. Jeffery (1922) obtained the time variation of $\phi_J$ and $\theta_J$ that is given by

$$\tan(\phi_J) = A_e \tan\left(2\pi \frac{t}{T} + \tau\right), \tag{5.3 a}$$

$$\tan(\theta_J) = \frac{A_e C}{\sqrt{A_e^2 \cos^2(\phi_J) + \sin^2(\phi_J)}}, \tag{5.3 b}$$

where $\phi_J \in [0, 2\pi)$, $\theta_J \in [0, \pi]$, $C$ is the orbit constant, $\tau$ is the phase angle, $T = 2\pi(A_e + A_e^{-1})$ is the period of rotation of $\boldsymbol{p}$ and $A_e$ is the effective aspect ratio of the particle defined as $A_e = \sqrt{(1 + \lambda_J)/(1 - \lambda_J)}$. A thin cylinder ($A \gg 1$) spends most of its time such that $\boldsymbol{p}$ is near the flow-vorticity plane ($\phi_J \rightarrow \pi/2$). The rate of change of $\psi_J$ is given by

$$\frac{d\psi_J}{dt} = -\frac{1}{2}\lambda_J \cos(\theta_J) \cos(2\phi_J) \tag{5.3 c}$$

According to equation (5.3 c), $\psi_J$ changes over a timescale of $O((1 + (A_e C)^2)^{0.5})$ which varies with the orbit constant, $C$, and contrasts from the fixed $O(T)$ time scale over which $\theta_J$ and $\phi_J$ change. Therefore, there are uncountably infinite orbits where $\psi_J$ rotates quasi-periodically while only countably infinite orbits where $\psi_J$ has a period that is a multiple of $T$. The quasi-periodic rotation of $\psi_J$ demonstrates that the $(\psi_J, \phi_J)$ space is filled completely over time as seen in figures 7 (b) and (c). Discrete peaks in the frequency spectrum obtained by the fast Fourier transformation $\widehat{\psi}_J$ is used to establish the quasi-periodic nature of the system as shown in figure 7 (d).

The average velocity of the particle is zero because of the symmetry of the orbits relative to the SSF. The quasiperiodic translation is quantified using the root mean square (r.m.s) velocity of the particle given by

$$\langle (\boldsymbol{U} \cdot \boldsymbol{\zeta})^2 \rangle^{1/2} = \lim_{N \to \infty} \frac{1}{\sqrt{NT}} \left( \sqrt{\int_0^{NT} (\boldsymbol{U} \cdot \boldsymbol{\zeta})^2 dt} \right), \tag{5.4}$$

where $\boldsymbol{\zeta} \in \left\{ \widehat{\boldsymbol{\Omega}}_\infty, \widehat{\boldsymbol{\Gamma}}_\infty \right\}$ and $N$ should be sufficiently large such that the results are invariant on increasing $N$. The mean-square velocity can also be obtained by averaging in the $(\psi_J, \phi_J)$ space given by

$$\langle (\boldsymbol{U} \cdot \boldsymbol{\zeta})^2 \rangle = \int_0^{2\pi} d\phi_J \left( \frac{1}{2\pi} \right) \left( \frac{A_e(1-\lambda)}{1+\lambda \cos(2\phi_J)} \right) \int_0^{2\pi} d\psi_J \frac{1}{2\pi} (\boldsymbol{U} \cdot \boldsymbol{\zeta})^2. \tag{5.5}$$

The r.m.s. velocity in the gradient $(\widehat{\boldsymbol{\Gamma}}_\infty)$ and vorticity $(\widehat{\boldsymbol{\Omega}}_\infty)$ direction for $\alpha_3 = 0.1$ are shown in figure 8 (a) and (b) respectively as a function of $C$ for varying particle aspect ratios. The r.m.s. velocities computed from equation (5.5) match the values obtained from time averaging over 500 tumbling events reaffirming the quasi-periodic nature of the system. For $C \to 0$ the particle is symmetric about the flow-gradient plane and therefore cannot translate in the vorticity direction as evident in figure 8 (b). In this orbit $\widehat{\boldsymbol{X}} = \widehat{\boldsymbol{U}}_\infty$ and $\widehat{\boldsymbol{Y}} = \widehat{\boldsymbol{\Gamma}}_\infty$ thereby leading to the highest value of $U_y$ among the orbits as per equation (5.2). Therefore, $\left\langle \left( \boldsymbol{U} \cdot \widehat{\boldsymbol{\Gamma}}_\infty \right)^2 \right\rangle$ is highest when $C = 0$ and monotonically decreases with increasing $C$ as seen in figure 8 (a). A peak is observed in $\left\langle \left( \boldsymbol{U} \cdot \widehat{\boldsymbol{\Omega}}_\infty \right)^2 \right\rangle$ for $C \sim O(1/A)$ before the value plateaus for large $C \gg 1$.

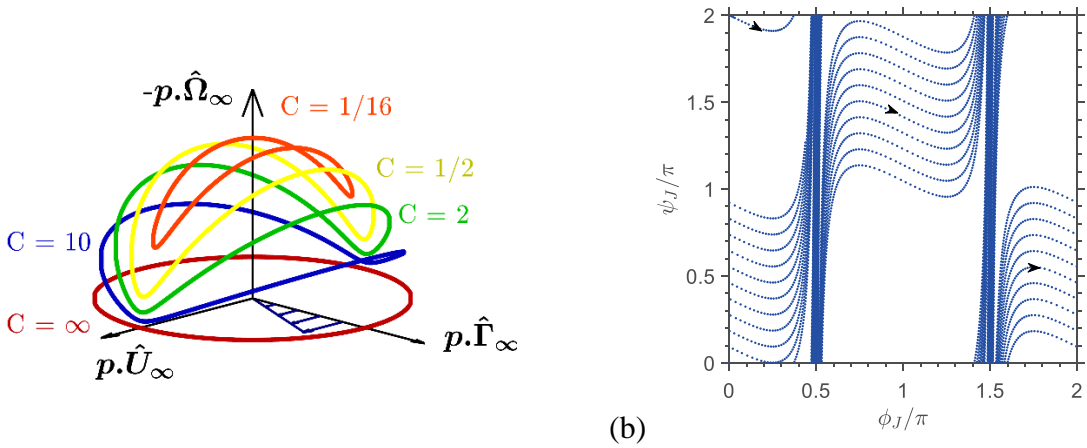

(a)                              (b)

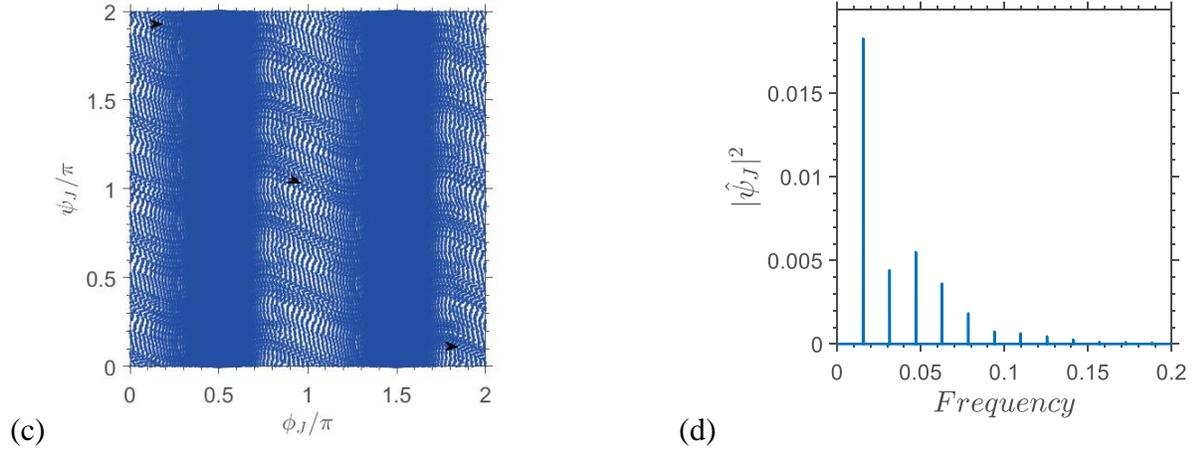

Figure 7. Trajectory of the orientation of the particle. (a) Orientational trajectory of $\boldsymbol{p}$ ($\lambda_J = 0.98$) for various initial conditions specified by the orbit constant, C. Change in $(\psi, \phi)$ during (b) 10 and (c) $10^3$ tumbling events respectively for a cylinder with $A = 20$ and $C = 0.1$. (d) Frequency spectrum obtained from the Fast Fourier transformation of $\psi_J(t)$ for $C = 0.1$ and $A = 20$.

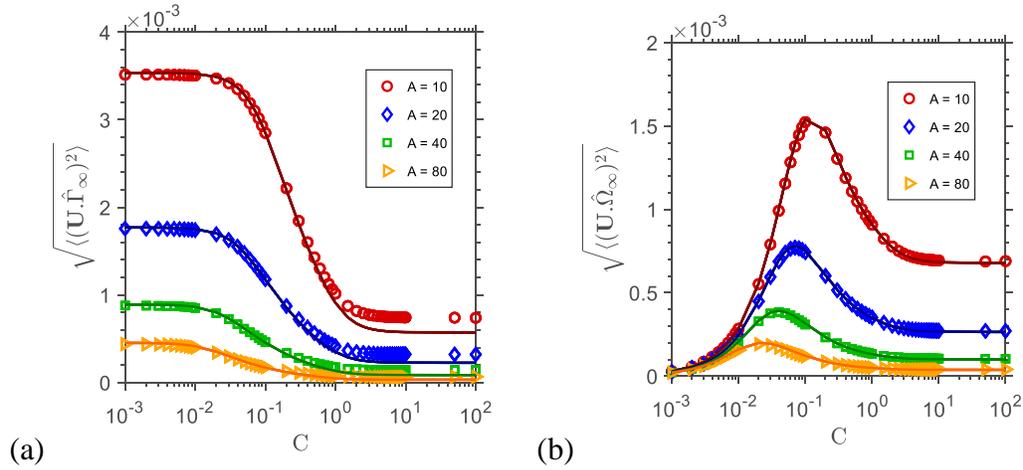

Figure 8. Particle r.m.s. velocity variation with particle aspect ratio, $A$, and orbit constant, $C$ for $\alpha_3 = 0.1$. Root mean square velocity of the particle in (a) the gradient direction and (b) the vorticity direction averaged over 500 tumbling events. Solid lines are phase-space integration using equation (5.5), while the symbols are the results of the time integration (equation 5.4).

These qualitatively new results predicting cross-stream drift due to cross-sectional asymmetries can be compared to the drift velocities observed for curved fibers (Wang et al. 2012) and screw like particles (Kim and Rae 1991). Curved fibers with an aspect ratio 20 and a curvature of unity migrate in the gradient direction with an average velocity of $1.7 \times 10^{-3}$ (Wang et al. 2012). The drift velocity of screw shaped particles is shown to be $O(10^{-4})$, where the length and time are non-dimensionalized using the length of the screw along its axis and the inverse of the shear rate respectively (The diameter of the screw was 1, the diameter of the filament was 0.1, and the screw had two turns.) Both these values which are numerically comparable to the values in figure 8 (a) suggest that the translation of slender bodies caused by cross-sectional modifications can have a similar magnitude to the effects of the shape of the centerline.

### 5.3. Diffusive translation of particles

Chaotic rotation and diffusive translation of a straight particle with a cross-section that is a combination of an ellipse and a third Fourier mode perturbation to a circle is demonstrated. The cross-section is given by $\rho_s = (l_1^2 \cos^2(\theta - \theta_{02}) + l_2^2 \sin^2(\theta - \theta_{02}))^{0.5} + a\alpha_3 \cos(3\theta)$, where $l_1, l_2$ are lengths of the semi-major axes of an ellipse such that $l_1 > l_2$, $a$ is the radius of the equivalent circle of the ellipse with semi-axes $l_1$, $l_2$ obtained from the analysis in section (3.3) and $\alpha_3$ is the amplitude of the third Fourier mode perturbation to the equivalent circle. The cross-section is chosen to be longitudinally elliptic (i.e., $a/a_0 = l_1/l_{1,0} = l_2/l_{2,0} = (1 - s^2)^{0.5}$) with $\theta_{02} = 0$ as such particles are known to rotate chaotically when $\alpha_3 = 0$ (Yarin et al. 1997). The rotational motion of such particles can be described using Jeffery's (1922) equations of motion since the third Fourier mode perturbation does not alter $\boldsymbol{\omega}$ as shown earlier. Yarin et al. (1997) demonstrated chaotic rotation of a particle with $\alpha_3 = 0$, $l_{1,0} = 2/10$ and $l_{2,0} = 1/10$ (an ellipsoid). Such a particle with a finite $\alpha_3 = 0.2$ would translate diffusively in addition to rotating chaotically. This particle has $a_0 = 1.5/10$, $\boldsymbol{K}$ is represented in terms of $\bar{\alpha}_2 = 0.33$, $\theta_{02}$ and $K_{zz} = 0$; and $\boldsymbol{L}$ is represented in terms of $\bar{\alpha}_3 = 0.20$, $\theta_{03} = 0$ and $L_z = 0$, according to equation (3.18). The values $\bar{\alpha}_2 = 0.33$, $\bar{\alpha}_3 = \alpha_3$ and $\theta_{03} = 0$ are accurate within 7% error for $\forall \, \alpha_3 \leq 0.2$ and arbitrary $\theta_{02}$.

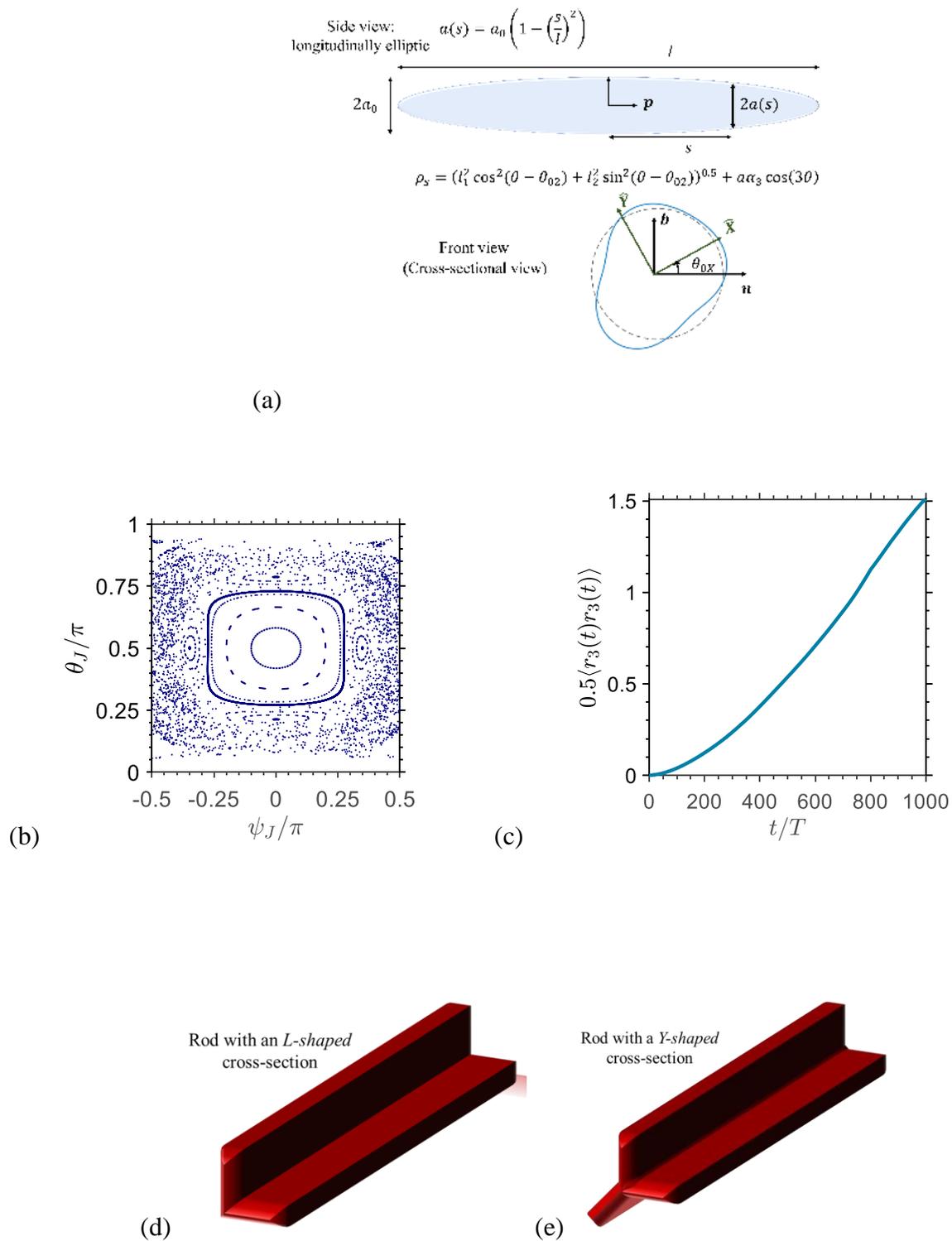

(a)

(b)

(c)

(d)

(e)

Figure 9. Shape and chaotic dynamics of particles. (a) Representative shape that can rotate and translate chaotically in a SSF. (b) Poincare map sampled when $\phi_J$ is a multiple of $\pi$ for the shape

in (a) with $\alpha_3 = 0.1$, $l_{1,0} = 2/10$ and $l_{2,0} = 1/10$ that shows the chaotic sea. The closed loops represent trajectories in which $(\phi_J, \theta_J, \psi_J)$ change quasi-periodically. A detailed Poincare map can be seen in figure 6 (a) of Yarin et al. (1997). (c) Variation of $\langle r_3(t)r_3(t)\rangle$ with the time, suggests a diffusive behavior of the particle position. (d) A straight rod with an *L-shaped*, which can rotate chaotically and migrate diffusively. (e) A straight rod with a *Y-shaped* cross-section that can rotate periodically and translate quasi-periodically. These particles can be fabricated using photolithography (Foulds and Parameswaran 2006) or 3D printing (Raney and Lewis 2015).

The particle motion is obtained using the numerical procedure mentioned in Yarin et al. (1997) for ellipsoids. The Poincare map sampled when $\phi_J$ is a multiple of $\pi$ shows the chaotic sea as seen in figure 9 (b). A particle whose orientation starts within the sea will span it after enough time. The simulation was carried out for a time of $10^4 T$, where $T$ is the time period of rotation of a spheroid of the same aspect ratio, $A = 1/a_0$. The diffusivity in the gradient ($D_{33}$) and vorticity ($D_{22}$) direction, obtained from the position of the particle, is given by

$$D_{ij} = \tfrac{1}{2}\lim_{t\to\infty}\tfrac{d}{dt}\langle r_i(t)r_j(t)\rangle. \qquad (5.6)$$

The variation of $\langle r_3 r_3\rangle$ with time shown in figure 9 (c) suggests a ballistic motion at short times and a diffusive behavior for long times with a diffusivity of $4 \times 10^{-5}$. The diffusivity in the vorticity direction $D_{22} = 6 \times 10^{-8}$ is of much smaller magnitude. The current case is particularly interesting as the particle is self-dispersive at zero Reynolds number without Brownian diffusion or inter-particle interactions. This gradient diffusivity is numerically comparable to the gradient diffusivity of a fiber of the same length and aspect ratio $A$ due to interparticle interactions when the dimensional particle number density $n^* L^{*3} \approx 0.37$, where $L^*$ is the dimensional length of the particle (Rahnama et al. 1993; Lopez and Graham 2007). Similarly, the gradient diffusivity of curved fibers with aspect ratios of approximately 20 is $O(10^{-5})$ (Wang et al. 2014), again of a magnitude similar to the influence of the cross-section.

The orientational dynamics in this section illustrated for straight cylinders which are longitudinally elliptic can be extended to other straight bodies with tapered ends using results from a complimentary study of Cox (1971). Cox (1971) obtained the $O(1/(\ln(A)\,A^2))$ torque acting on a stationary body with tapered ends and a circular cross-section held stationary in the flow-vorticity plane of the simple shear flow. This torque can be matched with the torque required to rotate the particle in a quiescent fluid to obtain $\boldsymbol{\omega}$. On applying a regular perturbation of the inner solution of Cox (1971) one can extend his result to a slightly non-circular cross-section. The details can be found in Cox (1971) which is discussed in section (S. 3) of the supplementary material.

For a cylinder with blunt edges, which is a more practical case, the analysis of Cox (1971) breaks down because the ends of the cylinder generate an $O(1/A^2)$ torque on the particle. The torque on a stationary cylinder with blunt ends is equal to $2\boldsymbol{p} \times \boldsymbol{F}_{end}$, where $\boldsymbol{F}_{end}$ is the force acting on an end of the particle. For a general shaped cross-section, one can find a second-order tensor $\boldsymbol{A}$, such that $\boldsymbol{F}_{end} = \boldsymbol{A} \cdot \boldsymbol{g}$. $\boldsymbol{A}$ can be derived by taking 3 random orientations and finding $\boldsymbol{F}_{end}$ from experiments or numerical solutions of the Stokes equations. The part of $\boldsymbol{\omega} \cdot (\boldsymbol{I} - \boldsymbol{pp})$ driven by the straining part of the SSF, can be obtained by equating $2\boldsymbol{p} \times \boldsymbol{F}_{end}$ to the torque required to rotate a particle in a quiescent fluid $8\pi\boldsymbol{\omega} \cdot (\boldsymbol{I} - \boldsymbol{pp})/(3\ln(A))$. The longitudinal component of the angular velocity $\boldsymbol{\omega} \cdot \boldsymbol{p}$ can be obtained by matching the torque due to $\boldsymbol{g} \cdot \boldsymbol{p}$ to $4\pi\boldsymbol{\omega} \cdot \boldsymbol{pp} \int ds\, a^2$ (Cox 1971). This can allow us to model the rotational dynamics of straight particles with a general cross-section and blunt edges using the SBT formulation. This calculation of the orientation dynamics of straight particles is important in predicting the structure and rheology of fiber suspensions which could be useful in paper-manufacturing research.

A straight rod with an *L-shaped* cross-section shown in figure 9 (d) has a finite value of $\bar{\alpha}_2$ and $\bar{\alpha}_3$ and thereby rotates and translates chaotically (a non-zero $\bar{\alpha}_2$ is similar to an equivalent elliptic cross-section). A straight rod with a *Y-shaped* cross-section shown in figure 9 (e) has a non-zero value of $\bar{\alpha}_3$ while $\bar{\alpha}_2 = 0$ and therefore rotates periodically and translates quasi-periodically. These rods could be fabricated via multi-step photolithography (Foulds and Parameswaran, 2006) or 3D printing (Raney and Lewis 2015) opening a pathway to experimentally verify our results and observe interesting dynamics. Einarsson et al. (2016) measured the rotational motion of non-axisymmetric particles formed by connecting multiple micro-rods, which can simulate particles with two or three lobed cross-sections. The results presented here demonstrate the nature of cross-sectional shapes that can be used to control the rotational and translational dynamics of straight particles in a SSF.

## 6. Motion of rings in a simple shear flow (SSF)

In this section, the SBT is used to predict the dynamics of rings with non-circular cross-sections and the results are verified using boundary element method (BEM) calculations. The influence of the cross-sectional geometry on the rotational and translational motion of rings is established using analytical expressions. Rings with cross-sections that have $\bar{\alpha}_2 \neq 0$ rotate and translate periodically with no net migration over time if the contribution to the third Fourier mode, $\bar{\alpha}_3$, is below a critical value that depends on the aspect ratio. On the other hand, rings with a $\bar{\alpha}_3 \neq 0$ can attain an equilibrium orientation and can drift indefinitely in the gradient direction if the aspect ratio is above a critical value that depends on $\bar{\alpha}_3$. Such rings can self-align without application of external forces or torques, thereby creating a highly anisotropic structure that can be contrasted with the dispersed particle orientation in a suspension of rotating particles. Here, the length is non-dimensionalized using the radius of the ring R. The shear rate of the SSF and the viscosity of the fluid are used to non-dimensionalize other quantities of interest such as the force per unit length and the linear and angular velocity of the particle.

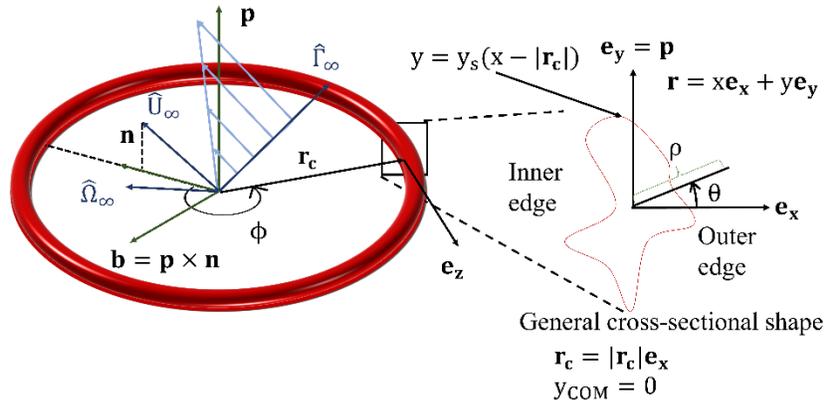

Figure 10. Geometry of a ring and the four coordinate systems, namely the global coordinate system $(\widehat{\boldsymbol{U}}_\infty, \widehat{\boldsymbol{\Omega}}_\infty, \widehat{\boldsymbol{\Gamma}}_\infty)$, the particle coordinate system $(\boldsymbol{n}, \boldsymbol{b}, \boldsymbol{p})$, the local Cartesian coordinate system $(\boldsymbol{e_x}, \boldsymbol{e_y}, \boldsymbol{e_z})$ and the local polar coordinate system $(\rho, \theta)$ in the cross-section plane.

## 6.1    Coordinate system

The global coordinate system is defined along the flow ($\widehat{\boldsymbol{U}}_\infty$), vorticity ($\widehat{\boldsymbol{\Omega}}_\infty$) and gradient ($\widehat{\boldsymbol{\Gamma}}_\infty$) directions of a SSF as shown in figure 10. The coordinate system relative to the particle is defined along the axis of symmetry, $\boldsymbol{p}$, a vector $\boldsymbol{n} = (\boldsymbol{I} - \boldsymbol{pp}) \cdot \widehat{\boldsymbol{U}}_\infty$ and a vector $\boldsymbol{b} = \boldsymbol{p} \times \boldsymbol{n}$. The local coordinate system ($\boldsymbol{e_x}, \boldsymbol{e_y}, \boldsymbol{e_z}$), is defined such that $\boldsymbol{e_y} = \boldsymbol{p}$, $\boldsymbol{e_z}$ is tangent to the centerline of the ring cross-section, $\boldsymbol{r_c}$, and $\boldsymbol{e_x} = \boldsymbol{e_y} \times \boldsymbol{e_z}$ (normal to the centerline of the ring cross-section). Here, $\boldsymbol{r_c}$ is the separation vector of the centerline of the ring cross-section relative to the center of mass (COM) of the ring. The center of the cross-section is chosen to coincide with the apparent hydrodynamic center of the cross-section described in section (3.3). The azimuthal angle $\phi$, measured from $\boldsymbol{n}$ in the plane of the ring (i.e., $\boldsymbol{n} - \boldsymbol{b}$ plane), determines the position along the centerline of the ring cross-section. A local polar coordinate system ($\rho$-$\theta$) is defined in the $\boldsymbol{e_x} - \boldsymbol{e_y}$ plane, where $\theta$ is measured from $\boldsymbol{e_x}$ and $\rho$ is the normal distance from $\boldsymbol{r_c}$. The aspect ratio of the ring is $A = 1/a$, where $a$ is the radius of the unperturbed circular cross-section.

### 6.2 Dynamics of rings with slightly non-circular cross-sections

The functional form of the angular velocity ($\boldsymbol{\omega}$) of a ring in an unbounded linear flow field using the linearity of Stokes flow is given by

$$\boldsymbol{\omega} = 0.5\boldsymbol{\epsilon} : \boldsymbol{W}_\infty + \lambda \, \boldsymbol{p} \times (\boldsymbol{E}_\infty \cdot \boldsymbol{p}), \tag{6.1}$$

where $\boldsymbol{p}$ is the particle orientation, $\boldsymbol{W}_\infty = 0.5(\nabla \boldsymbol{u}_\infty - (\nabla \boldsymbol{u}_\infty)^T)$ is the vorticity tensor, $\boldsymbol{E}_\infty = 0.5(\nabla \boldsymbol{u}_\infty + (\nabla \boldsymbol{u}_\infty)^T)$ is the strain rate and $\lambda$ is the rotation parameter (Bretherton (1962) and Jeffery (1922)). The drift velocity ($\boldsymbol{U} - \boldsymbol{u}_\infty \, (\boldsymbol{r_{COM}})$) of the particle relative to the fluid velocity at its centre-of-mass (COM) can be written as

$$\boldsymbol{U} - \boldsymbol{u}_\infty(\boldsymbol{r_{COM}}) = \eta_1 \boldsymbol{E}_\infty \cdot \boldsymbol{p} + \eta_2 (\boldsymbol{ppp} : \boldsymbol{E}_\infty), \tag{6.2}$$

where $\eta_1$ and $\eta_2$ are the translation parameters that depend only on the particle geometry (Brenner 1964; Singh, et al. 2013). The values of the dynamic parameters of the particle, $\lambda, \eta_1$ and $\eta_2$ can be obtained by applying the force-free ($\int \boldsymbol{f}_{net}(\phi) R d\phi = \boldsymbol{0}$) and torque-free ($\int \left( (\boldsymbol{r} - \boldsymbol{r}_{COM}) \times \boldsymbol{f}_{net}(\phi) + \boldsymbol{g} \right) R d\phi = \boldsymbol{0}$) conditions on the particle, where $\boldsymbol{g} = \int ds_c (\boldsymbol{r} - \boldsymbol{r}_c) \times (\boldsymbol{\sigma} \cdot \widetilde{\boldsymbol{n}}) + (\boldsymbol{r}_c - \boldsymbol{r}_{COM}) \times \int ds_c \, \cos(\theta) / A (\boldsymbol{\sigma} \cdot \widetilde{\boldsymbol{n}})$, $\boldsymbol{\sigma}$ is the stress tensor obtained from the solution of the 2-D Stokes equations and $\nabla^2 u_z = 0$ with $\boldsymbol{u} = \boldsymbol{u}_\infty$ on the outer boundary and no-slip on the particle surface, $\widetilde{\boldsymbol{n}}$ is the unit normal to the surface of the particle and $ds_c$ is the elemental length along the cross-sectional contour (see supplementary S.4 for details). For rings, equation (3.12) can be solved analytically using elliptic integrals to obtain $\boldsymbol{f}_{net}$ with errors of $O(1/A^2)$ (see supplementary material S.4 for details). $\lambda, \eta_1$ and $\eta_2$ are given by

$$\lambda = -1 - \frac{\alpha_3 \cos(3\theta_{03})}{A} + \frac{(\ln(8A) - 1.5)}{A^2} C_\lambda + O\left(\frac{\alpha_3^2}{A}\right) + O\left(\frac{\alpha_2^2}{A}\right) + O\left(\frac{\alpha_2 \alpha_3}{A}\right), \tag{6.3}$$

$$\eta_1 = -\frac{2}{3} \frac{\alpha_3 \sin(3\theta_{03})}{A} \frac{(\ln(8A) - K_{zz}/2 - 3)}{\ln(8A) - K_{zz}/3 - 17/6} + O\left(\frac{\alpha_2^2}{A}\right) + O\left(\frac{\alpha_3^2}{A}\right) + O\left(\frac{\alpha_2 \alpha_3}{A}\right), \tag{6.4}$$

$$\eta_2 = \frac{17}{12} \frac{\alpha_3 \sin(3\theta_{03})}{A} \frac{(\ln(8A) - 7K_{zz}/17 - 99/34)}{\ln(8A) - K_{zz}/3 - 17/6} + \frac{\alpha_2 \sin(2\theta_{02})}{4\ln(8A) - 10} + O\left(\frac{\alpha_2^2}{\ln(8A)}\right) + O\left(\frac{\alpha_3^2}{A}\right) + O\left(\frac{\alpha_2 \alpha_3}{\ln(8A)}\right), \tag{6.5}$$

where $C_\lambda = A^2 \left( \int d\phi \, \boldsymbol{g} \cdot \widehat{\boldsymbol{\Omega}}_\infty \right) / (2\pi^2) \sim O(1)$, $\boldsymbol{g}$ being the torque per unit length computed when $\boldsymbol{p} = \hat{\boldsymbol{\Gamma}}_\infty$. Here, $K_{zz}$ is the effect of the cross-sectional shape on the longitudinal velocity field ($\boldsymbol{e_z}$) as described in section (3.3) and is zero at $O(\alpha_2)$ and $O(\alpha_3)$. The part of the value of $\lambda$ equal to $-1 - \alpha_3 \cos(3\theta_{03})/A$, is the contribution due to $\boldsymbol{f}_{net}$ and the $O(\ln(8A)/A^2)$ term is the contribution due to $\boldsymbol{g}$. Both of these terms are crucial for particles that can self-align in a simple shear flow for which $\lambda + 1$ crosses zero. For a general cross-sectional shape equations (6.3) - (6.5) can be used if $(\alpha_2, \alpha_3)$ are replaced with $(\bar{\alpha}_2, \bar{\alpha}_3)$, where $(\bar{\alpha}_2, K_{zz}, \bar{\alpha}_3)$ are related to $\boldsymbol{K}$ and $\boldsymbol{L}$ according to equation (3.18). For a circular cross-section $C_\lambda = 1.5$ and this value maintains great accuracy when $\bar{\alpha}_2$ and $\bar{\alpha}_3$ are small.

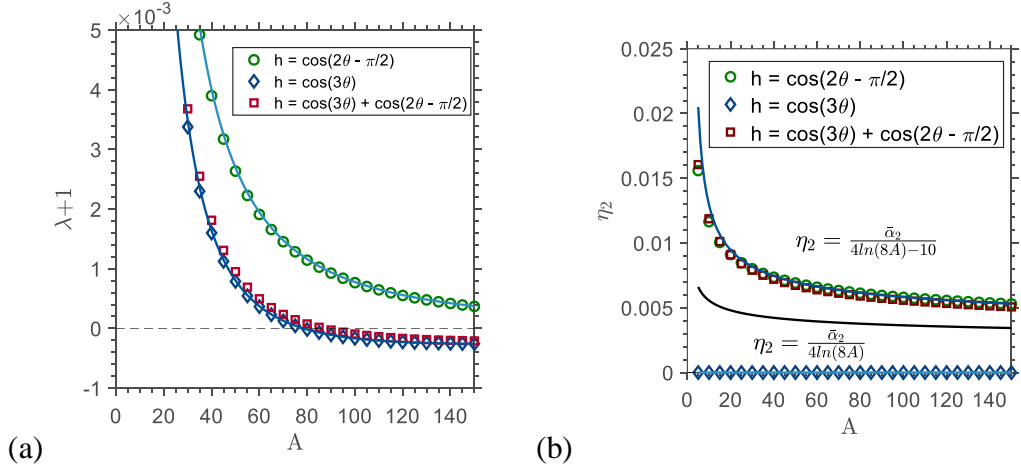

Figure 11. Comparison of SBT with the numerical results obtained from BEM calculations (Borker et al. 2018). Variation of (a) $\lambda$ and (b) $\eta_2$, with aspect ratio, $A$ (solid lines), and verification with the boundary element method calculations (symbols) for shapes shown in figure 1 (b)-(d). The cross-sectional shape is defined as $\rho_c = a(1 + 0.1h)$.

The values of $\lambda$ and $\eta_2$ obtained from equations (6.3), (6.5) using $(\bar{\alpha}_2, \bar{\alpha}_3, K_{zz}) = (0.0975, 0.0945, 0.0)$ (as obtained using the numerical calculation in section 3.3 instead of $(\alpha_2, \alpha_3) = (0.1, 0.1)$), compare well with the numerical values obtained from BEM described in Borker et al. (2018) as shown in figure 11. Figure 11 (a) shows that $\lambda$ for shapes S-1 and S-3 are nearly identical and close to the SBT prediction. This result confirms the prediction of SBT that at linear order in $\alpha$, $\lambda$ is only affected by the perturbation to a circle given by $h(\theta) = \cos(3\theta - 3\theta_{03})$. The dynamical parameters $\lambda$ and $\eta_2$ are accurately predicted because the integral equation (3.12) is solved with algebraic errors of $O(1/A^2)$. For comparison, $\eta_2$ obtained from the leading-order solution $\bar{\alpha}_2/(4\ln(8A))$ is only qualitatively accurate as seen in figure 11 (b).

### 6.3 Dynamics of rings that can self-align in a SSF

The translational and rotational motion of rings with an arbitrary cross-sectional shape can be specified using equations (6.1) – (6.5) and requires only the solution of a 2-D Stokes flow problem mentioned in section (3.3). The time evolution of the orientation and position of a ring has four qualitatively different states: (i) continuous periodic tumbling without cross-stream translation ($\bar{\alpha}_3 < C_\lambda(\ln(8A) - 1.5)/A$, $\sin(2\theta_{02}) = 0$, $\sin(3\theta_{03}) = 0$), (ii) continuous periodic tumbling with periodic translation ($\bar{\alpha}_3 < C_\lambda(\ln(8A) - 1.5)/A$ & ($\bar{\alpha}_2 \sin(2\theta_{02}) \neq 0$ or $\bar{\alpha}_3 \sin(3\theta_{03}) \neq 0$)), (iii) equilibrium orientation without cross-stream translation ($\bar{\alpha}_3 \geq C_\lambda(\ln(8A) - 1.5)/A$, $\bar{\alpha}_3 \neq 0$, $\sin(2\theta_{02}) = 0$, $\sin(3\theta_{03}) = 0$), and (iv) equilibrium orientation with a net translation in the gradient direction of the SSF ($\bar{\alpha}_3 \geq C_\lambda(\ln(8A) - 1.5)/A$, $\bar{\alpha}_3 \neq 0$ & ($\bar{\alpha}_2 \sin(2\theta_{02}) \neq 0$ or $\sin(3\theta_{03}) \neq 0$)). Cases (i) and (ii) can be studied using traditional SBT formulations of Cox (1970) and Batchelor (1970) respectively and are not treated here. The qualitative nature of ring dynamics in cases (iii) and (iv) cannot be captured using any previous SBT formulations to the best of our knowledge.

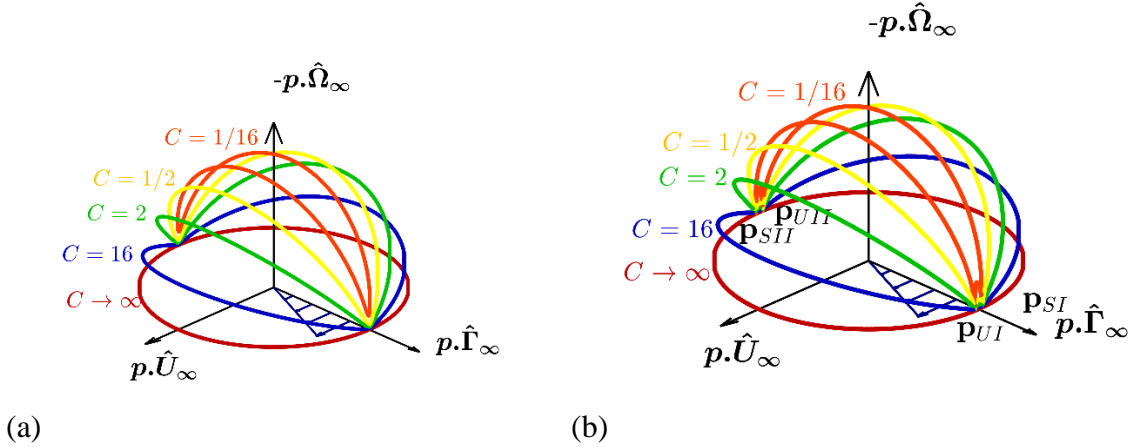

Figure 12. Trajectories traced by the orientation vector of a ring ($\boldsymbol{p}$) (a) with a circular cross-section and an aspect ratio $A = R/a = 100$ and (b) with the three-lobed cross section shown in figure 1 (c) with $\bar{\alpha}_3 = 0.3$, $\theta_{03} = 0$ and $A = R/a = 100$. This high value of $\alpha_3$ was chosen to yield a visually apparent difference between the stable ($\boldsymbol{p}_{SI}, \boldsymbol{p}_{SII}$) and unstable ($\boldsymbol{p}_{UI}, \boldsymbol{p}_{UII}$) nodes.

A ring with the cross-section shown in figure 1 (c) (S-II shape) shows qualitatively different rotational dynamics from a ring with a circular or S-I shaped cross-section. Such rings attain an equilibrium orientation, as shown in figure 12 (b), instead of rotating continuously in Jeffery orbits as shown in figure 12 (a). The particle orientation, $\boldsymbol{p}$, aligns along one of the two stable nodes ($\boldsymbol{p}_{SI}, \boldsymbol{p}_{SII}$) in the flow-gradient plane which are very close to the gradient direction, as shown in figure 12 (b). This was first shown by Singh et al. (2013) for rings with S-II shaped cross-sections with $\theta_{03} = 0$. Rings with an S-II shaped cross-section can in general align for a non-zero $\theta_{03}$ as evident from equation (6.3), which can be confirmed based on the physical explanation of alignment given in Borker et al. (2018).

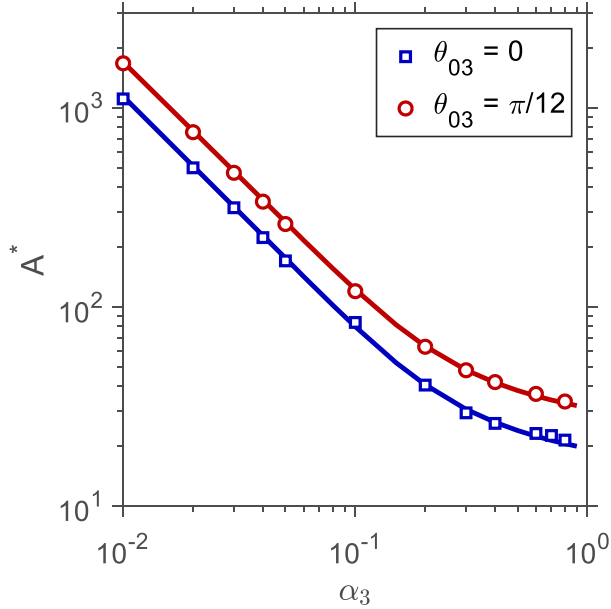

Figure 13. Variation of the critical aspect ratio, $A^*$, vs $\alpha_3$ from SBT (lines) compared with BEM results of Borker et al. (2018) (symbols) for $\theta_{03} = 0$ and $\theta_{03} = \pi/12$. $\bar{\alpha}_3$ lies in the range $0.01 \leq \bar{\alpha}_3 \leq 0.27$ for $0.01 \leq \alpha_3 \leq 0.9$.

The critical aspect ratio, $A^*$, defined as the value of A for which $\lambda = -1$, is a quantity of interest as it is the lowest aspect ratio at which a ring with a given cross-sectional shape can align. A lower value of $A^*$ would also mean that the ring will be less prone to bending and buckling, thus improving the structural integrity of the particle. Previously, a large number of computationally expensive boundary element method calculations would have been required to compute $A^*$. However, the current theory can be used to calculate $A^*$ from equation (6.6) when $\lambda = -1$ and is the solution to the equation given by

$$\frac{1}{A^{*2}}\left(A^* \bar{\alpha}_3 \cos(3\theta_{03}) - C_\lambda(\ln(8A^*) - 1.5)\right) = 0, \qquad (6.6)$$

where $\bar{\alpha}_3$ is obtained from the solution of the 2-D Stokes flow problem mentioned in section (3.3). Equation (6.6) only requires the solution of two 2-D Stokes flow problems and a solution to a Laplace's equation to obtain $\bar{\alpha}_3$ and $C_\lambda$. Figure 13 compares the prediction of $A^*$ from equation (6.6) with the BEM predictions for rings with a cross-section given by $\rho = a(1 + \alpha_3 \cos(3\theta - 3\theta_{03}))$. The accurate prediction of A* suggests that the current SBT framework can be used as a computationally inexpensive alternative to search for shapes that can self-align at the least $A^*$. Furthermore, equation (6.3) suggests that cross-sections with $\theta_{03} = 0$ (i.e. a fore-aft symmetric shape) and a large value of $\bar{\alpha}_3$, should have a smaller $A^*$ than fore-aft asymmetric shapes. The Y-shaped cross-section in Borker et al. (2018) which has the lowest reported value of $A^* = 8.9$ has features similar to the 3-lobed cross-section with $\theta_{03} = 0$.

Rings with S-II shaped cross-sections can migrate across streamlines in the gradient direction for non-zero values of $\theta_{03}$. These migrating rings could be deposited by flowing the suspension of particles along a surface allowing one to control the roughness or scratch resistance of the underlying surface (Isla et al. 2003). This drift velocity is given by $\eta_2(\boldsymbol{p_s} \cdot \widehat{\boldsymbol{U}}_\infty)(\boldsymbol{p_s} \cdot \widehat{\boldsymbol{\Gamma}}_\infty)^2 \approx \eta_2(0.5 |1 + \lambda|)^{0.5}$. The $O(\bar{\alpha}_2 \sin(2\theta_{02}) / \ln(A))$ drift due to the second Fourier mode perturbation is $O(A/\ln(A))$ larger than the drift due to the third Fourier mode perturbation. Therefore, shapes with $\theta_{03} \to 0$, which increases $|\lambda + 1|$, and $\theta_{02} \to \pi/4$, which maximizes $\eta_2$, should generate the highest drift velocities. The cross-section which led to the highest drift shown in figure 9 (a) of Borker et al. (2018) also has a clear resemblance to a shape which is a combination of the second and third Fourier mode with $\theta_{02} = \pi/4$.

Rings with the cross-sections shown in figure 14 (a) or (b) can align in a SSF at relatively low aspect ratios (Borker et al. 2018). These rings are of practical interest due to the ease of fabrication using multi-step photolithography (Foulds and Parameswaran, 2006) or optofluidic fabrication (Paulson, Di Carlo and Chung 2015), which can allow for testing the rheology of a suspension of such particles. Here, SBT is utilized to predict the dynamics of individual particles and the results are compared with the numerical predictions obtained using the boundary element method (BEM) detailed in Borker et al. (2018). COMSOL, a finite element solver, was used to perform the 2-D Stokes flow calculation presented in section (3.3). The values of $\boldsymbol{K}$ and $\boldsymbol{L}$, which were estimated with an uncertainty below 0.1 % when the size of the outer boundary ($\rho_\infty$) was at least 50 times the cross-sectional dimension, are reported in table 1. Figure 14 (c) and (d) show the variation of $\lambda$ with $A$ obtained from the BEM calculation and the corresponding SBT prediction for rings with *T-shaped* and *L-shaped* cross-sections, respectively. The SBT precisely mimics the BEM results even at low aspect ratios near the critical aspect ratio, $A^*$, which is possible by using the numerical procedure in section (3.3) and solving the integral in equation (3.17) with errors of $O(1/A^2)$ using elliptic integrals (supplementary material S. 4). $\eta_1$ and $\eta_2$ are zero for a ring with a *T-shaped* cross-section due to mirror symmetry about a plane normal to $\boldsymbol{p}$. Figures 14 (e) and (f) show the accuracy of SBT to predict the variation of $\eta_1$ and $\eta_2$ with $A$ for rings with the *L-shaped* cross-section. The force per unit length $\boldsymbol{f}$ obtained from equation (3.17) is also in excellent agreement with the BEM results for both $L$ and $T$ *shaped* cross-sections for $A \gtrsim 10$ as shown in figure 14 (g)-(h) by the value of $\boldsymbol{f}$ at $\phi = 0.2\pi$, $\boldsymbol{p} \cdot \widehat{\boldsymbol{\Omega}}_\infty = 0$ and $\boldsymbol{p} \cdot \widehat{\boldsymbol{\Gamma}}_\infty = \cos(0.2\pi)$. The dependence of $\boldsymbol{f}$ on $\phi$ can be derived using the linearity of the governing equations (Borker et al. 2018) and the imposed boundary conditions and is presented in the supplementary material (S.4).

Table 1. $\boldsymbol{K}$ and $\boldsymbol{L}$ values for rings with *T-shaped* and *L-shaped* cross-sections represented in terms of $(\bar{\alpha}_2, \theta_{02}, K_{zz})$ and $(\bar{\alpha}_3, \theta_{03}, L_{zx}, L_{zy})$.

| *T-shaped* ring ($a = 1.055 \times l_T$) | $(\bar{\alpha}_2, \theta_{02}, K_{zz})$ | $(0.0345, 0, -0.05532)$ |
|---|---|---|
| | $(\bar{\alpha}_3, \theta_{03}, L_{zx}, L_{zy})$ | $(0.157, 0, -8.1 \times 10^{-4}, 0)$ |
| *L-shaped* ring ($a = 0.5446 \times l_L$) | $(\bar{\alpha}_2, \theta_{02}, K_{zz})$ | $(0.1534, -\pi/4, -0.05512)$ |
| | $(\bar{\alpha}_3, \theta_{03}, L_{zx}, L_{zy})$ | $(0.1366, -\pi/12, 1.4 \times 10^{-3}, 1.4 \times 10^{-3})$ |

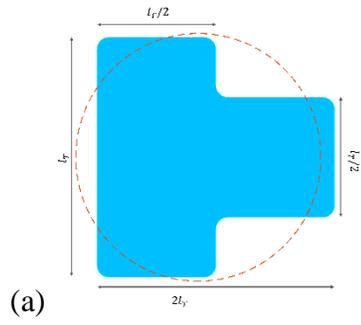

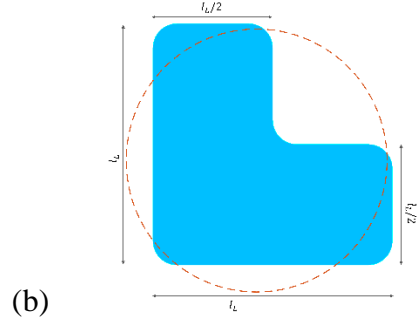

(a)

(b)

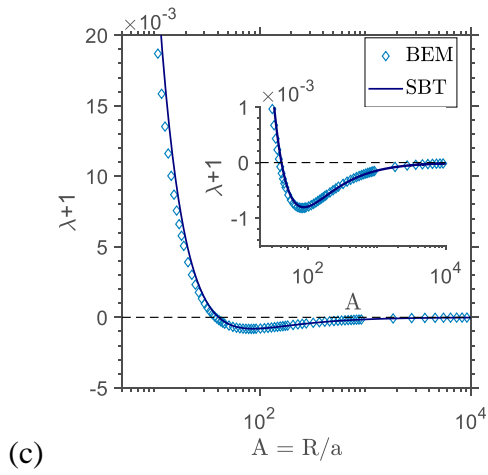

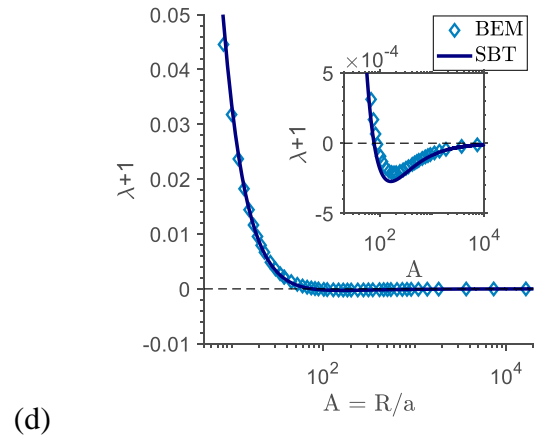

(c)

(d)

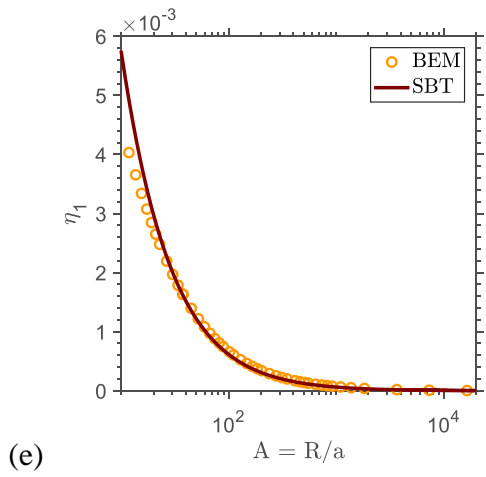

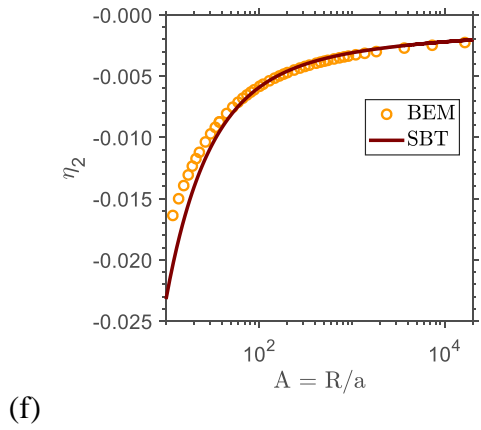

(e)

(f)

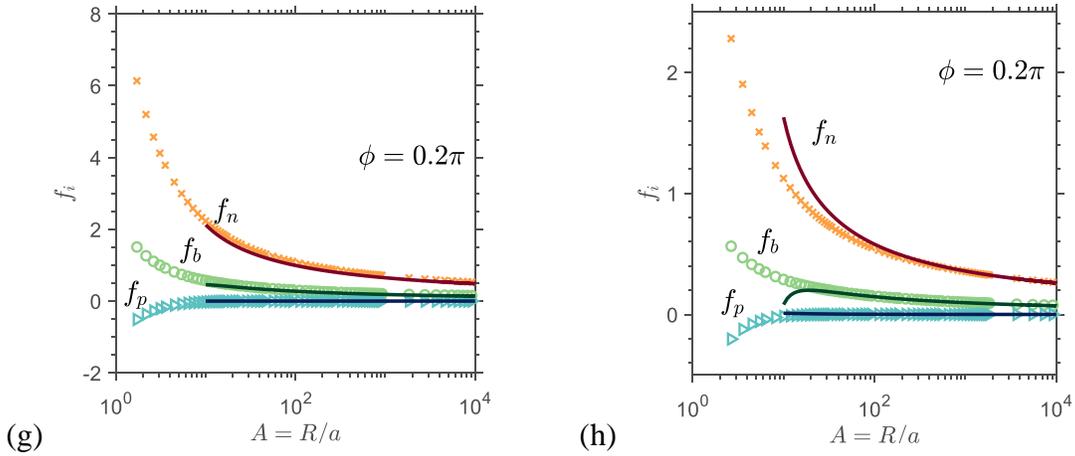

Figure 14. Application of current SBT to predict dynamics of rings with cross-sections which deviate significantly from a circle. (a) A "*T-shaped*" and (b) an "*L-shaped*" cross-section proposed in Borker et al. (2018). Rings with these cross-sections (shaded regions) can align in a SSF at finite aspect ratios. The equivalent circle obtained from the analysis in section (3.3) is shown using dotted lines. Comparison of $\lambda$ vs A variation from BEM for rings with (c) *T-shaped* and (d) *L-shaped* cross-sections with the SBT prediction from equation (6.3). Comparison of the variation of (e) $\eta_1$ and (f) $\eta_2$ with $A$ obtained from BEM calculations for rings with an *L-shaped* cross-section with the SBT predictions from equations (6.4) - (6.5). Force per unit length variation with $A$ at $\phi = 0.2\pi$, $\boldsymbol{p} \cdot \hat{\boldsymbol{\Omega}}_\infty = \boldsymbol{0}$ and $\boldsymbol{p} \cdot \hat{\boldsymbol{\Gamma}}_\infty = \cos(0.2\pi)$ for a ring with (g) *T-shaped* and (h) *L-shaped* cross-sections. Symbols are results from the BEM calculations, the solid lines are the corresponding values obtained from the current SBT formulation.

Figures 14 (g)-(h) suggest that our slender body theory accurately predicts $\boldsymbol{f}$ which can in turn be utilized to simulate hydrodynamic interactions between various rings, especially the ones that can self-align. This calculation is important for obtaining the rheology of a suspension of rings and also estimating the influence of surrounding particles on the self-aligning characteristics of such a ring. A ring aligned near the flow-vorticity plane has weak hydrodynamic interactions with other aligned rings and therefore the suspension should have a highly anisotropic microstructure, which can be expected to be retained at higher particle concentrations due to excluded volume interactions.

## 7. Conclusion

In this work, a slender body theory is developed for a thin, curved body with an arbitrary cross-section that allows one to solve for the velocity, pressure and force per unit length exerted by the particle on the fluid. The derivation is based on asymptotically matching the velocity field of an infinitely long cylinder in the inner region to the velocity field due to a line of forces in the outer region. Our theory accounts for the force per unit length associated with the gradient in the imposed fluid velocity which previously had not been embedded into SBT formulations. The features of the cross-sectional shape that display this qualitatively different force per unit length are described using regular perturbation of the inner solution. A cross-section that has two lobes or three lobes (figures 1 (b) and (c) respectively) will change the force per unit length by $O(\epsilon^2)$ or $O(\epsilon/A)$, respectively. Thought experiments that give physical insight into the special nature of the two and three lobed cross-sections were illustrated. A 2-D Stokes flow problem was formulated that can be numerically solved to extend our theory to arbitrary cross-sectional shapes which deviate significantly from a circle. Our analysis also captures the force per unit length driven by the relative velocity of the particle and the fluid for a non-circular cross-section, which was first derived by Batchelor (1970).

A slender cylinder that has a significant contribution to $\alpha_2$ and $\alpha_3$, e.g. an *L-shaped* cross-section (figure 8 d), rotates and translates chaotically, while a cylinder with a finite contribution to $\alpha_3$ along with $\alpha_2 = 0$, e.g. a *Y-shaped* cross-section (figure 8 e), rotates periodically and translates quasi-periodically. Our theory accurately predicted the resistance to translation and rotation for a triaxial ellipsoids even for high-aspect ratio cross-sections as shown by comparison with the exact results of Lamb (1932). In this case, the current method provides a computationally inexpensive alternative to other available approaches such as slender ribbon theory (Koens and Lauga, 2016) or boundary element method (Youngren & Acrivos, 1975; Kim & Karilla, 1991; Pozrikidis, 2002). The dynamics of rings with different cross-sectional shapes in a simple shear flow was used to further validate our theory by comparing the results with boundary element method calculations of Singh et al. (2013) and Borker et al. (2018). The perturbation analysis described in this paper could be extended to Stokes flow with fluid inertia (Khayat and Cox, 1989), potential flow (Lighthill 1960, 1971) and heat transfer (Beckers et al. 2015) to find the impact of the gradients in the respective background fields. The solution of the respective two-dimensional problem, similar to the problem described in section (3.3), could be used to extend the result to a general cross-sectional shape.

The advancement in nano-fabrication shows promise of utilizing micro and nanoscale objects with high-aspect ratio appendages to aid in targeted drug delivery, material assembly (Sacanna et al 2013) or water treatment (Gao and Wang 2014; Soler and Sanchez 2014). These slender micromachines would be subject to velocity gradients and our current work can be utilized to simulate their dynamics and thereby learn about optimal propulsion mechanism under a background flow field.

Our theory can also be utilized to understand slender particle dynamics in various linear flow fields. The results of section (5) suggest that the effects of the cross-section can have magnitudes similar to the effects of the curvature of the centerline of curved slender bodies. The current theory can be used to study the motion of straight particles in a simple shear flow (SSF), which has a rich dynamical structure. A sampling of such results were presented in section (5) for straight cylinders with mirror symmetry about the longitudinal direction where the $O(\alpha_3/A)$ force per unit length induced a net translation. The $O(\alpha_3/A)$ force per unit length, being proportional to $a$, will also induce an $O(\alpha_3/A)$ angular velocity to any particle that lacks mirror symmetry about the longitudinal direction. This $O(\alpha_3/A)$ angular velocity, which is important when the particle is near the flow-vorticity plane, has a much stronger scaling than the $O(\ln{(A)}/A^2)$ contribution from the dipole per unit length. Such particles can also translate with an $O(1)$ velocity arising from asymmetry along the longitudinal direction leading to velocities an order of magnitude larger than the ones presented in section (5). Furthermore, such asymmetric particles that have an additional contribution to $\alpha_2$ could also rotate and translate chaotically. The motion of straight cylinders can be explored using the current SBT formalism along with the solution of Cox (1971). Our work gives insight into the geometry of cross-sections that are important and the tools to explore the motion of such slender shapes in a shear flow.

The force per unit length acting on rings with non-circular cross-section, presented in section 6, can be used to simulate hydrodynamic interactions between multiple rings to obtain the structure and rheology of a suspension of rings. The particular case of interest is obtaining the rheology of rings that can attain an equilibrium orientation in a SSF, which has never been explored. A suspension of such aligned rings has the possibility of attaining high degrees of anisotropy due to alignment of all particles in the same orientation, which could be useful to manufacture highly anisotropic materials. Hydrodynamic interactions between rings can be captured by using equation (3.12) or (3.17) with the velocity disturbance produced by other rings included in the $\boldsymbol{u}_\infty$ term and solving for the force per unit length up to $O(1/\ln(2A))$.

**Acknowledgements**

This research was supported by the National Science Foundation Grants CBET-1435013 and CBET-1803156.

**References**

Acrivos, A, Shaqfeh, E. S. G. (1988). The effective thermal conductivity and elongational viscosity of a nondilute suspension of aligned slender rods. *Phys. Fluids* (1958–1988) 31, 1841–1844. (doi:10.1063/1.866681)

Anczurowski, E., and Mason, S. G., (1968) "Particle motions in sheared suspension. XXIV. Rotations of rigid spheroids and cylinders," *Trans. Soc. Rheol.* 12, 209–215

Batchelor, G. K. (1970). Slender-body theory for particles of arbitrary cross-section in Stokes flow. *J. Fluid Mech.*, 44(3), 419-440. doi:10.1017/S002211207000191X

Batchelor, G. K. (1954). The skin friction on infinite cylinders moving parallel to their length. *The Quarterly Journal of Mechanics and Applied Mathematics*, 7(2), 179-192.

Beckers, K. F., Koch, D. L., & Tester, J. W. (2015). Slender-body theory for transient heat conduction: theoretical basis, numerical implementation and case studies. *Proc. R. Soc. A* 471, 2184, 20150494.

Berg H C and Anderson R A 1973 Bacteria swim by rotating their flagellar filaments Nature 245 380–2.

Borker, N. S., Stroock, A. D., and Koch, D. L., Controlling rotation and migration of rings in a simple shear flow through geometric modifications, *J. Fluid Mech.* 840, 379-407. doi:10.1017/jfm.2018.20

Bray, D 2000 Cell Movements (New York: Garland)

Brennen, C and Winet, H 1977 Fluid mechanics of propulsion by cilia and flagella *Annu. Rev. Fluid Mech.* 9 339–98

Brenner, H. 1964 The Stokes resistance of an arbitrary particle. 3. shear fields. *Chem. Engng Sci.,* 19, 631–651

Bretherton, F. 1962 The motion of rigid particles in a shear flow at low Reynolds number. J. Fluid Mech. 14 (2), 284–304.

Chen, HS, Acrivos, A. (1976) On the effective thermal conductivity of dilute suspensions containing highly conducting slender inclusions. *Proc. R. Soc. Lond. A* 349, 261–276. (doi:10.1098/rspa.1976.0072)

Cox, R. G. (1970). The motion of long slender bodies in a viscous fluid Part 1. General theory. *Journal of Fluid Mechanics*, 44(4), 791-810. doi:10.1017/S002211207000215X

Cox, R. G. (1971). The motion of long slender bodies in a viscous fluid. Part 2. Shear flow. *Journal of Fluid Mechanics*, 45 (4), 625-657. (doi:10.1017/S0022112071000259)

Einarsson, J., Mihiretie, B. M., Laas A., Ankardal, S., Angilella, J. R., Hanstorp D. & B. Mehlig (2016) Tumbling of asymmetric microrods in a microchannel flow. *Physics of Fluid*, **28**, 013302. (https://doi.org/10.1063/1.4938239)

Foulds, I. G., & Parameswaran, M. (2006). A planar self-sacrificial multilayer SU-8-based MEMS process utilizing a UV-blocking layer for the creation of freely moving parts. *J. Micromech. and Microeng.*, 16 (10), 2109-2115. http://doi.org/10.1088/0960-1317/16/10/026

Fredrickson, G. H., Shaqfeh, E. S. G. (1989) Heat and mass transport in composites of aligned slender fibers. *Phys. Fluids A, Fluid Dyn*. (1989–1993) 1, 3–20. (doi:10.1063/1.857546)

Guasto, Jeffrey S., Roberto Rusconi, and Roman Stocker (2012) Fluid mechanics of planktonic microorganisms. Annual Review of Fluid Mechanics 44, 373-400. (https://doi.org/10.1146/annurev-fluid-120710-101156)

Gao, W., & Wang, J. (2014). The environmental impact of micro/nanomachines: a review. Acs Nano, 8(4), 3170-3180. (https://doi.org/10.1021/nn500077a)

Isla, A., Brostow, W., Bujard, B., Esteves, M., Rodriguez, J. R., Vargas, S. & Castano, V. M. 2003 Nanohybrid scratch resistant coatings for teeth and bone viscoelasticity manifested in tribology. Materials Research Innovations 7, 110–114.

Jeffery, G. B. (1922). The motion of ellipsoidal particles immersed in a viscous fluid. Proc. Roy. Soc. Lond. A,102 (715), 161-179. http://doi.org/10.1098/rspa.1922.0078

Johnson, R. E. (1980). An improved slender-body theory for Stokes flow. *Journal of Fluid Mechanics*, 99(2), 411-431. (https://doi.org/10.1017/S0022112080000687)

Johnson, R. E. & Brokaw, C. J. (1979) Flagellar hydrodynamics. A comparison between resistive-force theory and slender-body theory. *Biophys. J.* 25, 113–127. (https://doi:10.1016/S0006-3495(79)85281-9)

Jones, R. T. (1946) Properties of low-aspect-ratio pointed wings at speeds below and above the speed of sound. *Report no. 835. Washington, DC: National Advisory Committee for Aeronautics.*

Keller, J., & Rubinow, S. (1976). Slender-body theory for slow viscous flow. Journal of Fluid Mechanics, 75(4), 705-714. (https://doi:10.1017/S0022112076000475)

Khayat, R.E. and Cox, R.G., 1989. Inertia effects on the motion of long slender bodies. Journal of Fluid Mechanics, **209**,435-462.

Kim, Sangtae, and Karrila, Seppo J., 2013 Microhydrodynamics: principles and selected applications. Dover, 2013.

Kim, Y. J. and Rae, W. J. (1991). Separation of screw-sensed particles in a homogeneous shear field, *Int. J. Multiph. Flow*, 17(6), 717-744.(http://dx.doi.org/10.1016/0301-9322(91)90053-6)

Koens, L., & Lauga, E. (2016). Slender-ribbon theory. *Physics of Fluids*, 28(1), 013101.

Lamb, S. H., (1932). Hydrodynamics University Press.

Lighthill, M. J. 1960 Note on the swimming of slender fish. *J. Fluid Mech.,* 9, 305–317. (doi:10.1017/S0022112060001110)

Lighthill, M. J. 1971 Large-amplitude elongated-body theory of fish locomotion. Proc. R. Soc. Lond. B 179, 125–138. (https://doi:10.1098/rspb.1971.0085)

Lopez, M. & Graham, M. D. 2007 Shear-induced diffusion in dilute suspensions of spherical or non-spherical particles: Effects of irreversibility and symmetry breaking. *Phys. Fluids* 19, 073602.

Mackaplow, M. B, Shaqfeh, E. S. G. 1996 A numerical study of the rheological properties of suspensions of rigid, non-Brownian fibres. *J. Fluid Mech.*, 329, 155–186. (https://doi:10.1017/S0022112096008889)

Mackaplow, M. B, Shaqfeh, ESG. 1998 A numerical study of the sedimentation of fibre suspensions. *J. Fluid Mech.*, 376, 149–182. (https://doi:10.1017/S0022112098002663)

Mackaplow, M. B, Shaqfeh, E. S. G., Schiek, R. L. 1994 A numerical study of heat and mass transport in fibre suspensions. Proc. R. Soc. Lond. A 447, 77–110. (https://doi:10.1098/rspa.1994.0130)

Munk, M. M. 1924 The aerodynamic forces on airship hulls. Report no. 184. Washington, DC: *National Advisory Committee for Aeronautics*.

Newman, J. N. (1964) A slender-body theory for ship oscillations in waves. *J. Fluid Mech.* 18, 602–618. (https://doi:10.1017/S0022112064000441)

Newman, J. N. (1970) Applications of slender-body theory in ship hydrodynamics. *Annu. Rev. Fluid Mech.* 2, 67–94. (https://doi:10.1146/annurev.fl.02.010170.000435)

Paulsen, K. S., Di Carlo, D. and Chung, A. J. (2015). Optofluidic fabrication for 3D-shaped particles. *Nature Communications* 6, 6976. http://doi.org/10.1038/ncomms7976

Pozrikidis, C. 2002 A Practical Guide to Boundary Element Methods With the Software Library BEMLIB. Chapman and Hall/CRC

Rahnama, M., Koch, D. L., Iso, Y., & Cohen, C. (1993). Hydrodynamic, translational diffusion in fiber suspensions subject to simple shear flow. *Physics of Fluids A: Fluid Dynamics*, 5(4), 849-862. (https://doi.org/10.1063/1.858890)

Rahnama, M., Koch, D. L., & Shaqfeh, E. S. G.(1995). The effect of hydrodynamic interactions on the orientation distribution in a fiber suspension subject to simple shear flow. *Physics of Fluids*, **7**(3), 487-506.

Raney, J. R. & Lweis, J. A. 2015 Printing mesoscale architectures. *MRS Bull.* 40 (11), 943–950.

Rocha, A, Acrivos, A. (1973 a) On the effective thermal conductivity of dilute dispersions: general theory for inclusions of arbitrary shape. *Q. J. Mech. Appl. Math.*, **26**, 217–233. (https://doi:10.1093/qjmam/26.2.217)

Rocha, A, Acrivos, A. (1973 b) On the effective thermal conductivity of dilute dispersions: highly conducting inclusions of arbitrary shape. *Q. J. Mech. Appl. Math.* **26**, 441–455. (https://doi:10.1093/qjmam/26.4.441)

Sacanna, S, Korpics, M, Rodriguez, K, Colón-Meléndez, L, Kim, S-H, Pine, D. J., Yi, Gi-Ra, Shaping colloids for self-assembly (2013) *Nature Communications*, 4, 1688.

Shaqfeh, E. S. G. (1988) A nonlocal theory for the heat transport in composites containing highly conducting fibrous inclusions. *Phys. Fluids* (1958–1988) **31**, 2405–2425. (https://doi:10.1063/1.866594)

Singh, V., Koch, D. L., & Stroock, A. D., (2013). Rigid ring-shaped particles that align in simple shear flow. *Journal of Fluid Mechanics*, 722, 121-158. (doi:10.1017/jfm.2013.53)

Soler, L., & Sánchez, S. (2014). Catalytic nanomotors for environmental monitoring and water remediation. *Nanoscale*, 6(13), 7175-7182.

Suarez, S. S., & Pacey, A. A. (2006). Sperm transport in the female reproductive tract. *Human reproduction update*, 12(1), 23-37

Tekce, H. S., Kumlutas, D., & Tavman, I. H. (2007). Effect of Particle Shape on Thermal Conductivity of Copper Reinforced Polymer Composites. Journal of Reinforced Plastics and Composites, 26(1), 113–121. (https://doi.org/10.1177/0731684407072522)

Wang, J., Graham, M. D., Klingenberg, D. J. 2014 Shear-induced diffusion in dilute curved fiber suspensions in simple shear flow *Physics of Fluids* 26(3), 033301. (https://doi.org/10.1063/1.4867171)

Wang, J., Tozzi, E. J., Graham, M. D., & Klingenberg, D. J.2012 Flipping, scooping, and spinning: Drift of rigid curved nonchiral fibers in simple shear flow *Phys. Fluids* 24, 123304.

Yarin, A. L., Gottlieb, O., & Roisman, I. V. (1997). Chaotic rotation of triaxial ellipsoids in simple shear flow. *J. Fluid Mech.*, **340**, 83-100. ([https://doi.org/10.1017/S0022112097005260](https://doi.org/10.1017/S0022112097005260))

Youngren, G. K. & Acrivos, A. 1975 Stokes flow past a particle of arbitrary shape: a numerical method of solution. *J. Fluid Mech.* **69** (02), 377–403.